\documentclass[a4paper,11pt]{article}
\pdfoutput=1 

\usepackage{jinstpub} 
\usepackage{stackengine}
\usepackage{upgreek}
\usepackage{threeparttable}
\usepackage{cases}
\usepackage{adjustbox}
\usepackage{rotating}
\usepackage{lscape}
\usepackage{afterpage}

\usepackage{amsmath}
\usepackage{upgreek}
\usepackage{cases}
\usepackage{fixltx2e}
\usepackage{threeparttable}
\usepackage{hyperref}
\usepackage{subfig}
\usepackage{graphicx}

\title{A massively scalable Time-to-Digital Converter with a PLL-free calibration system in a commercial 130 nm process}

\author[a,b,1]{F. Martinelli\note{Corresponding author.}}
\author[c]{, P. Valerio}
\author[d]{, R. Cardarelli}
\author[b]{, E. Charbon}
\author[c]{, G. Iacobucci}
\author[a]{, M. Nessi}
\author[a,c]{and L. Paolozzi}

\affiliation[a]{The European Organization for Nuclear Research (CERN), \\Espl. des Particules 1, 1211 Meyrin, Switzerland}
\affiliation[b]{École polytechnique fédérale de Lausanne (EPFL), Advanced Quantum Architecture (AQUA) Laboratory\\Rue de la Maladière 71C, 2002 Neuchâtel, Switzerland}
\affiliation[c]{University of Geneva, Département de physique nucléaire et corpusculaire (DPNC)\\Quai Ernest-Ansermet 24, 1205 Geneva, Switzerland}
\affiliation[d]{Istituto Nazionale Fisica Nucleare (INFN), 00133 Rome, Italy}

\emailAdd{fulvio.martinelli@cern.ch}

\abstract{A 33.6 ps LSB Time-to-Digital converter was designed in 130 nm BiCMOS technology. The core of the converter is a differential 9-stage ring oscillator, based on a multi-path architecture. A novel version of this design is proposed, along with an analytical model of linearity. The model allowed us to understand the source of the performance superiority (in terms of linearity) of our design and to predict further improvements. The oscillator is integrated in a event-by-event self-calibration system that allows avoiding any PLL-based synchronization. For this reason and for the compactness and simplicity of the architecture, the proposed TDC is suitable for applications in which a large number of converters and a massive parallelization are required such as High-Energy Physics and medical imaging detector systems. A test chip for the TDC has been fabricated and tested. The TDC shows a DNL$\leq$1.3 LSB, an INL$\leq$2 LSB and a single-shot precision of 19.5 ps (0.58 LSB). The chip dissipates a power of 5.4 mW overall. }


\begin{document}
\maketitle
\flushbottom

\section{Introduction}
\label{sec:introduction}
Time-to-digital converters (TDCs) have a significant impact on the performance of timing detectors, whenever high resolution is sought. 
In medical imaging or High-Energy Physics (HEP) applications \cite{christiansen2004high} \cite{swann2004100}, the integration of a large number of TDCs in a single chip with a time resolution better than 100 ps is often required to improve the quality of image reconstruction. For this reason, a simple, compact, easily scalable, low-power design is crucial for this kind of applications.
The TDC architecture proposed in this paper was designed with the aim of obtaining a converter that is able to combine all the specifications that high-time resolution pixel detector requires. This converter is based on a free-running RO that is able to perform an event-by-event measurement of the oscillation frequency which will compensate for potential (or unavoidable) drifts. Thus, this architecture allows implementing a simple and compact solution avoiding the use of any PLL-based synchronization system. This approach was first investigated during the development of various chips for timing detectors, as the ones produced for a full-silicon Positron Emission Tomography (PET) scanner at the University of Geneva \cite{paolozzi2019characterization} \cite{GrantPET} and for the proposal of a new preshower system for the FASER experiment at CERN. As anticipated, detectors for HEP and medical imaging applications can guarantee better performance if the system is featuring a large number of TDCs with time-resolution in the order of tens of picoseconds \cite{kolanoski2020particle}. Indeed, detectors with a more precise time measurement system are able to perform a better image reconstruction of the particles that they need to sense. For instance, in many PET scanners, the Time-of-Flight information is fundamental to reduce the positional uncertainty of the annihilation points of the positrons produced in the body under exam \cite{moses2003time}.
\begin{figure}[!b]
\centering
\includegraphics[width=3.45in]{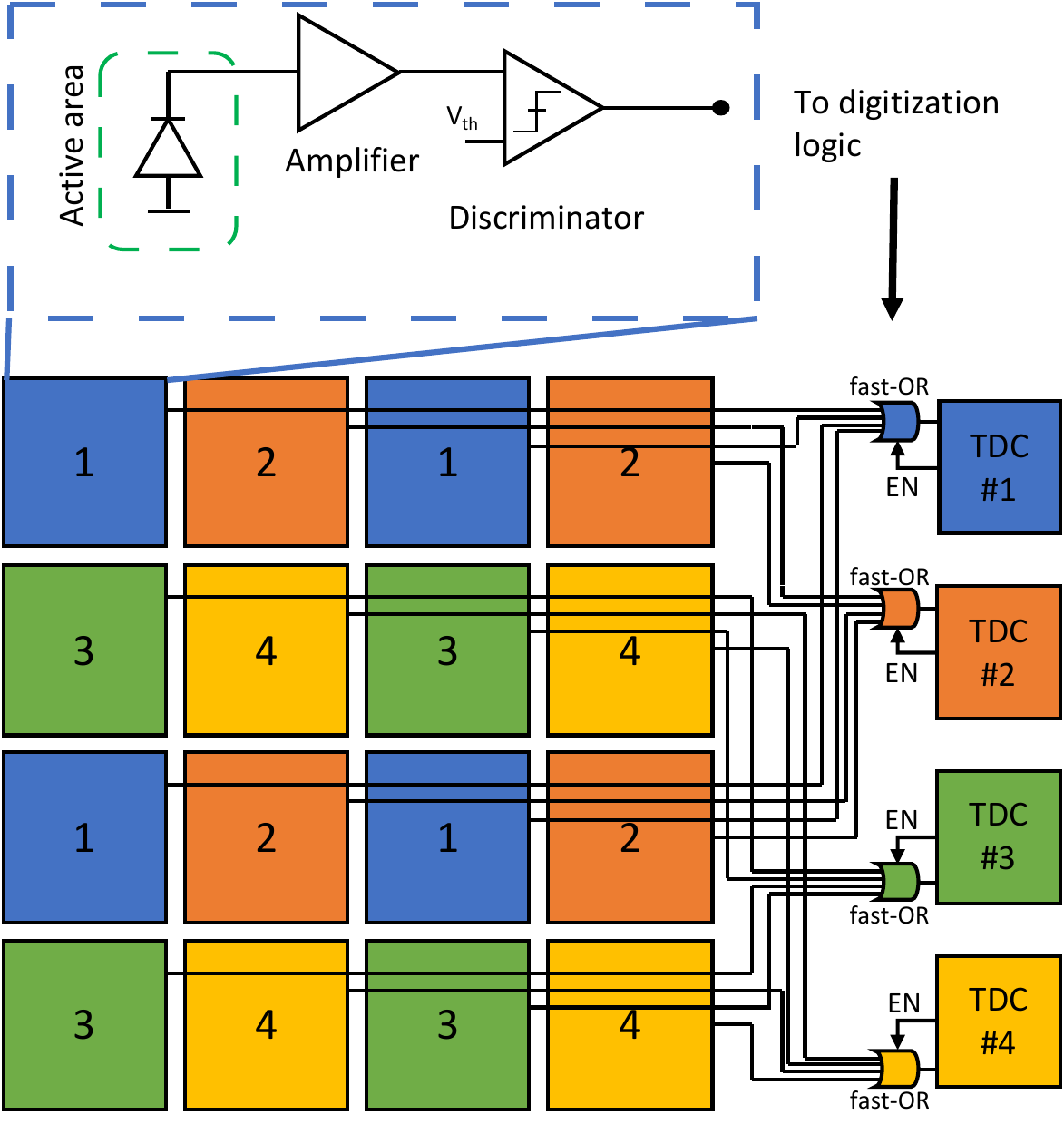}
\caption{Possible configuration of a 4 x 4 pixel matrix connected to 4 different TDC channels through fast-OR blocks. In this case active area refers to the sensitive region of the detecting system.}
\label{timing_detector_diagram}
\end{figure}
\\In a generic pixel detector with timing capabilities, having a structure in which each pixel is connected to its own TDC channel would be the ideal solution for efficiency purposes. Indeed, in this case, every portion of the matrix is independent of each other and the system will be able to store the timing information also in the case in which all the pixels are hit at the same time. However, especially for monolithic pixel detectors, this solution is difficult to implement for various reasons including area, complexity of the routing and power consumption. Hence, different design strategies need to be used, as the one illustrated in Figure \ref{timing_detector_diagram}. The matrix of the detector chip can be divided in sub-matrices: in the example of the figure, they are composed of 2 x 2 pixels and each of them is connected to a different TDC channel through the fast-OR blocks, together with the corresponding pixels of other sub-matrices. In this way, simultaneous hits on pixels of different channels (indicated with numbers from 1 to 4 in Figure \ref{timing_detector_diagram}) can be correctly detected. Having sub-matrices of pixels connected to separated converters avoids problems related to high cluster sizes because, in many detectors, the particles that need to be sensed can generate signals in groups of adjacent pixels \cite{jakubek2008pixel}.
The number of TDCs is chosen on the basis of the cluster size and the event rate, taking into account, as mentioned before, the power consumption and the area of the converter. If multiple hits occur on the same channel in a time window shorter than the dead time of the TDC, the converter, after the first one, will disable the fast-OR block in order to prevent other hits to interfere with the measurement. A possible improvement of this architecture is based on implementing a design that, in the multiple hits scenario, is able to store the position in the matrix of all the pixels that sensed an event after the first one without timing information. For all these reasons, the goal of the present work was to design a TDC characterized by a simple, compact and low-power design. Moreover, as will be shown in Section \ref{sec:arch}, the proposed converter is characterized by a PLL-less architecture, a useful solution to further reduce power consumption, complexity and area, integrating more TDC channels in a single chip.
\\The integration of the presented TDC inside a timing detector system requires a calibration process. Indeed, the difference among the delays of the ring oscillator and the counters used for the coarse component of the measurement can worsen the accuracy of the converter. In order to compensate this effect, a possible calibration approach is based on sending a periodic known event (synchronous with the reference clock) to the TDC. At this point, a set of offset parameters will be applied to the outputs of the system (given by Eq. \ref{2_eq:measurements_event_toa}-\ref{2_eq:measurements_event_cal} as it will be explained in Section \ref{sec:arch}) in order to minimize the standard deviation of the measured values.

\subsection{TDC basics and common architectures}
As introduced before, the development of a (tens of) picosecond-level resolution timing detector requires a TDC that is able to measure time with a precision in the same order of magnitude. Indeed, as explained in \cite{henzler2010time}, an ideal TDC is characterized by a quantization error (assuming a uniform distribution) with a standard deviation $\sigma_{q}$ proportional to the time of the Least Significant Bit (LSB) $T_{LSB}$
\begin{equation}
    \label{1_eq:sigma_ideal}
    \sigma_{q}=\frac{T_{LSB}}{\sqrt{12}}.
\end{equation}
This parameter is often indicated as resolution of the converter \cite{swann2004100}.
One of the traditional and most common approaches to design a TDC is based on using Ring Oscillators (ROs) \cite{nissinen2003cmos} \cite{krishna2020time} \cite{kim2018two}. Considering a certain time interval $T$, it is possible to measure a time difference by counting the number of cycles $N$ of the oscillator in the interval and sampling the RO at the edges of $T$, leading to
\begin{equation}
    \label{1_eq:measure_tdc_ro}
    T=N\cdot T_{RO}+T_{state}+\epsilon_{q},
\end{equation}
where $T_{RO}$ is the period of the RO, $T_{state}$ is the result of the sampling of the oscillator state which will produce the fine component of the measurement and $\epsilon_{q}$ is the quantization error. More recently, other architectures have been proposed. A possible implementation is presented in \cite{gebara20054} that shows an interpolative voltage-controlled oscillator (VCO). In this solution, the outputs of all the nodes of the structure are exploited to precharge further nodes in the oscillator resulting in an increase of the oscillation frequency. This implementation features a r.m.s. jitter value of 1.25 ps and a maximum frequency of 4.6 GHz in 180 nm CMOS technology and may be exploited for the design of both time digitizers and Phase-Locked Loop (PLL). A similar design approach has been adopted for the time conversion system integrated in the Blumino SiPM developed at EPFL \cite{muntean2020blumino}. The architecture proposed in the present paper features a similar mechanism to increase the oscillation frequency. Another solution that exploits a cyclic interpolation of switched-frequency RO allows measuring time intervals up to 375 $\upmu$s with a precision of 4.2 ps \cite{keranen2015wide}. 
\\In conventional RO-based architectures, the accuracy of the converter is given by the delay of the single cell of the oscillator $t_d$ \cite{henzler2010time}. In order to overcome this limitation, Vernier delay lines have often been used \cite{andersson2014vernier}: these solutions usually feature two delay lines with different stage delays $t_{d1}$ and $t_{d2}$, whereas the converter has a LSB equal to $\Delta=t_{d2}-t_{d1}$. However, the main limitation of this solution is represented by the measurement range of the converter that is given by $T_{max}=N\Delta$, where $N$ is the number of stages of the delay lines. For a certain value of $\Delta$, a wider range requires a larger $N$, thus resulting in a consequent increase of the power consumption. Various architectures can be implemented to overcome this trade-off such as cyclic Vernier lines to extend the maximum measurement time range, as the one presented in \cite{park2011cyclic}, or 2-D Vernier lines \cite{liscidini2009time}, which represent an efficient solution that allows obtaining $N$ quantization levels using only $\sqrt{N}$ stages. However, the complexity of these structures makes them unsuitable for the goals proposed before. 

\begin{figure*}[!t]
\centering
\subfloat[]{\includegraphics[width=2.5in]{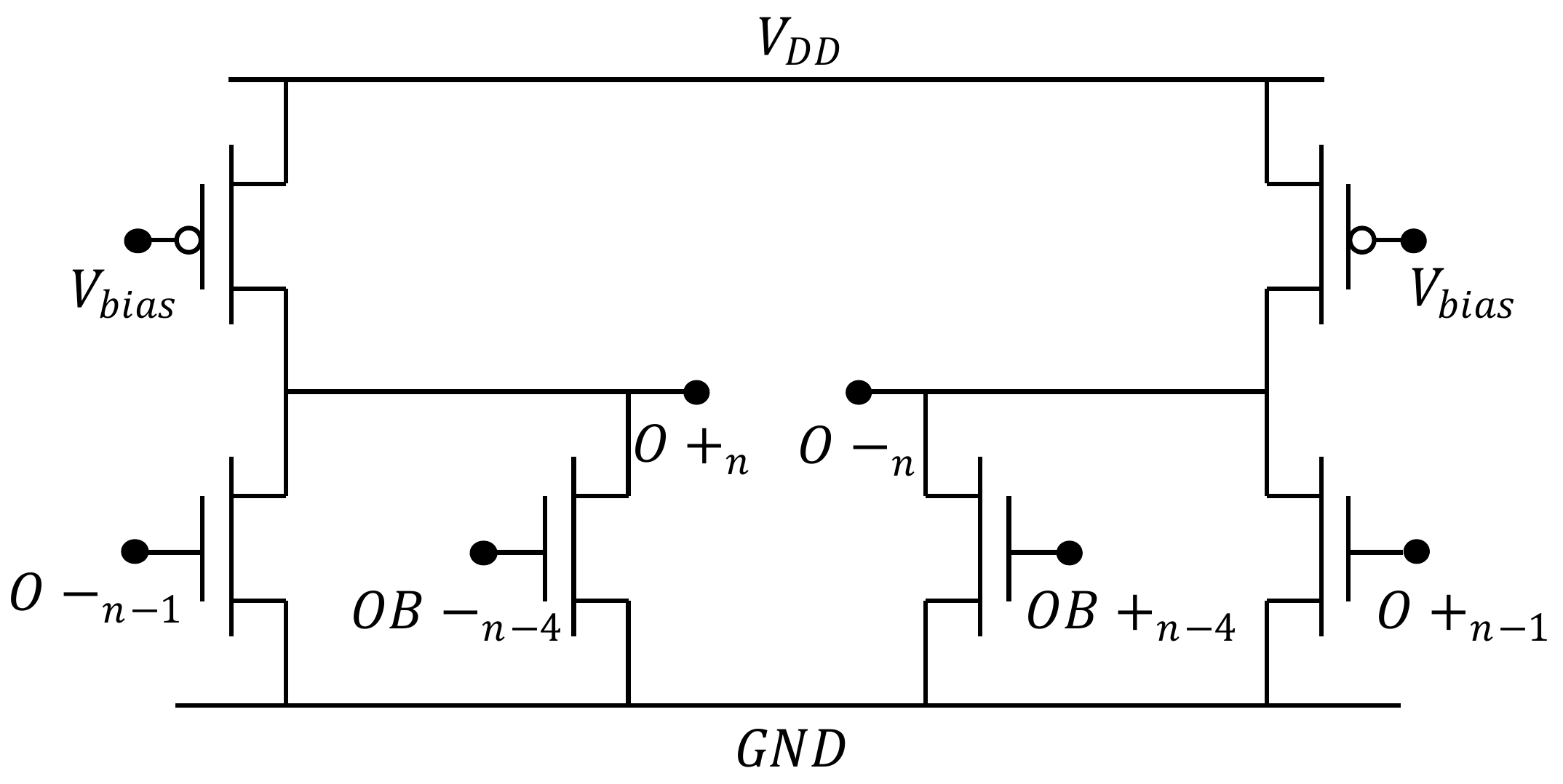}%
\label{delay_cell_buffer_first_case}}
\hfil
\subfloat[]{\includegraphics[width=2.1in]{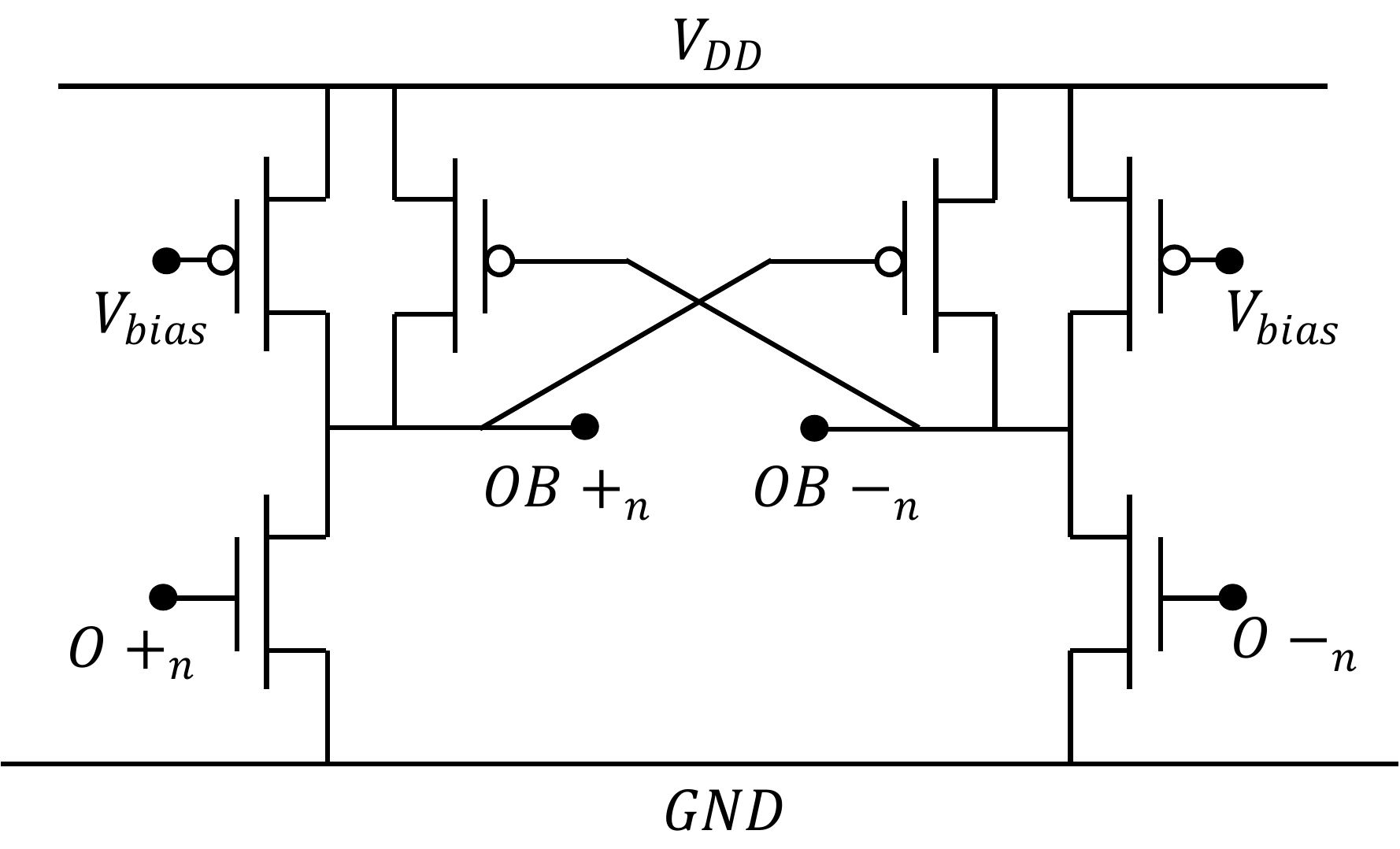}%
\label{delay_cell_buffer_second_case}}
\caption{Delay cell (a) and buffer (b) of the proposed RO.}
\label{delay_cell_buffer}
\end{figure*}
\begin{figure}[!t]
\centering
\includegraphics[width=4in]{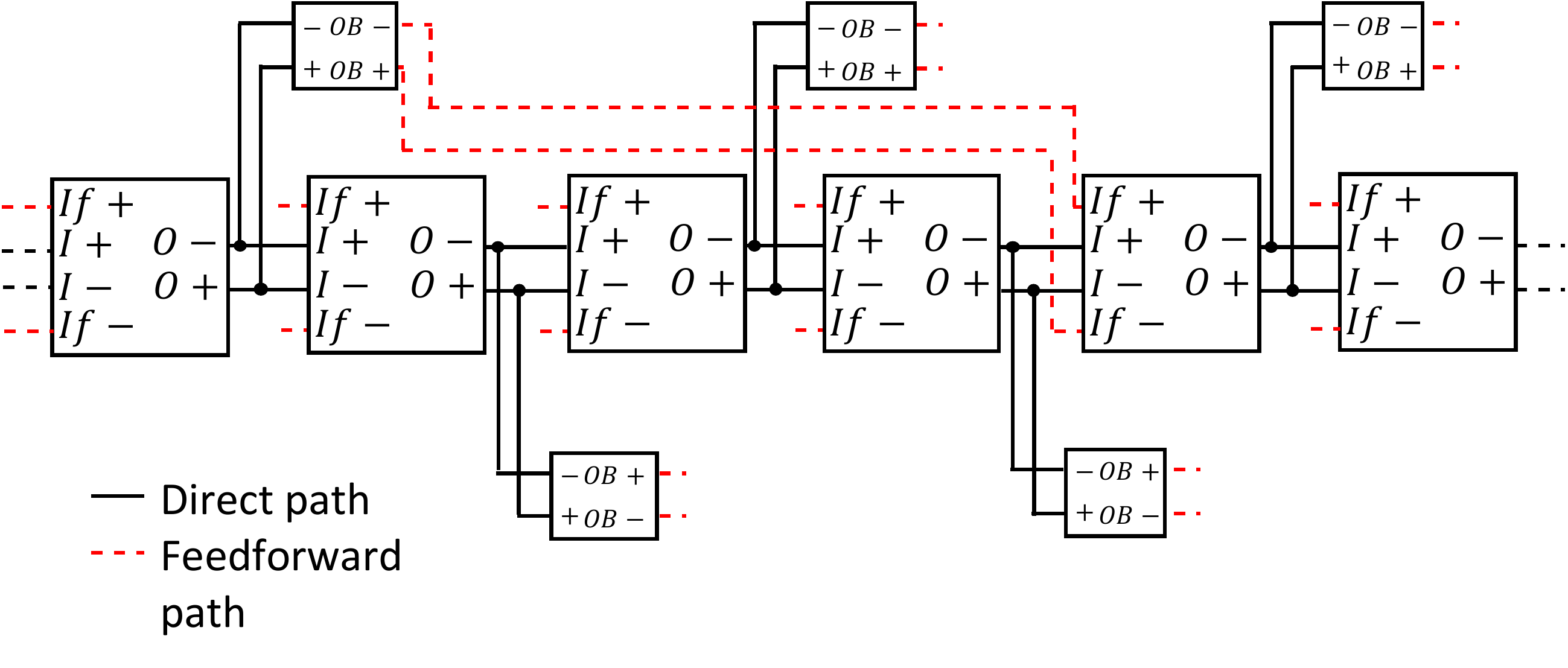}
\caption{Architecture of the proposed RO.}
\label{ring_oscillator}
\end{figure}
\begin{figure}[!t]
\centering
\includegraphics[width=3.5in]{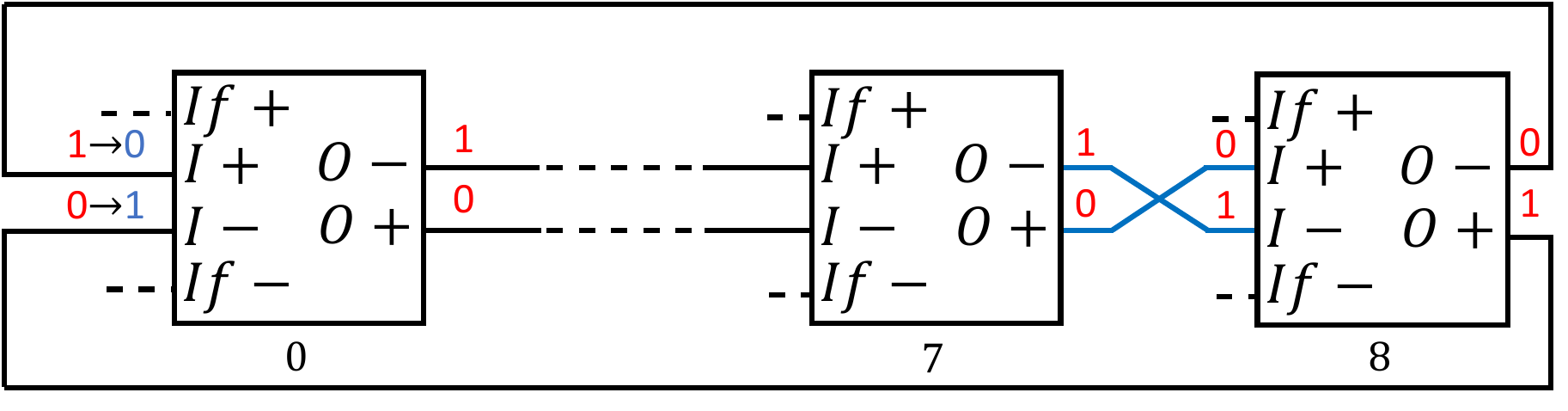}
\caption{Connections of the delay cells of the oscillator. Inverting the inputs of one of the stages allows the RO to properly oscillated satisfying the Barkhausen criterion.}
\label{ring_oscillator2}
\end{figure}

\section{Architecture}
\label{sec:arch}

The design process of the present TDC was not only focused on the implementation of a simple and compact architecture but also on the optimization of other fundamental parameters such as time-resolution and linearity that play a crucial role on the performance of timing detectors. This analysis was supported by analytical modeling and validated by extensive simulations. The proposed converter has been designed in 130 nm BiCMOS technology. The latter has been used by the group for the design of various pixel detectors and for their front-end systems. However, no bipolar transistor was used for the TDC and thus the analysis could be extended to a pure CMOS technology node. 

\subsection{Design}
\begin{figure}[!t]
\centering
\includegraphics[width=2.5in]{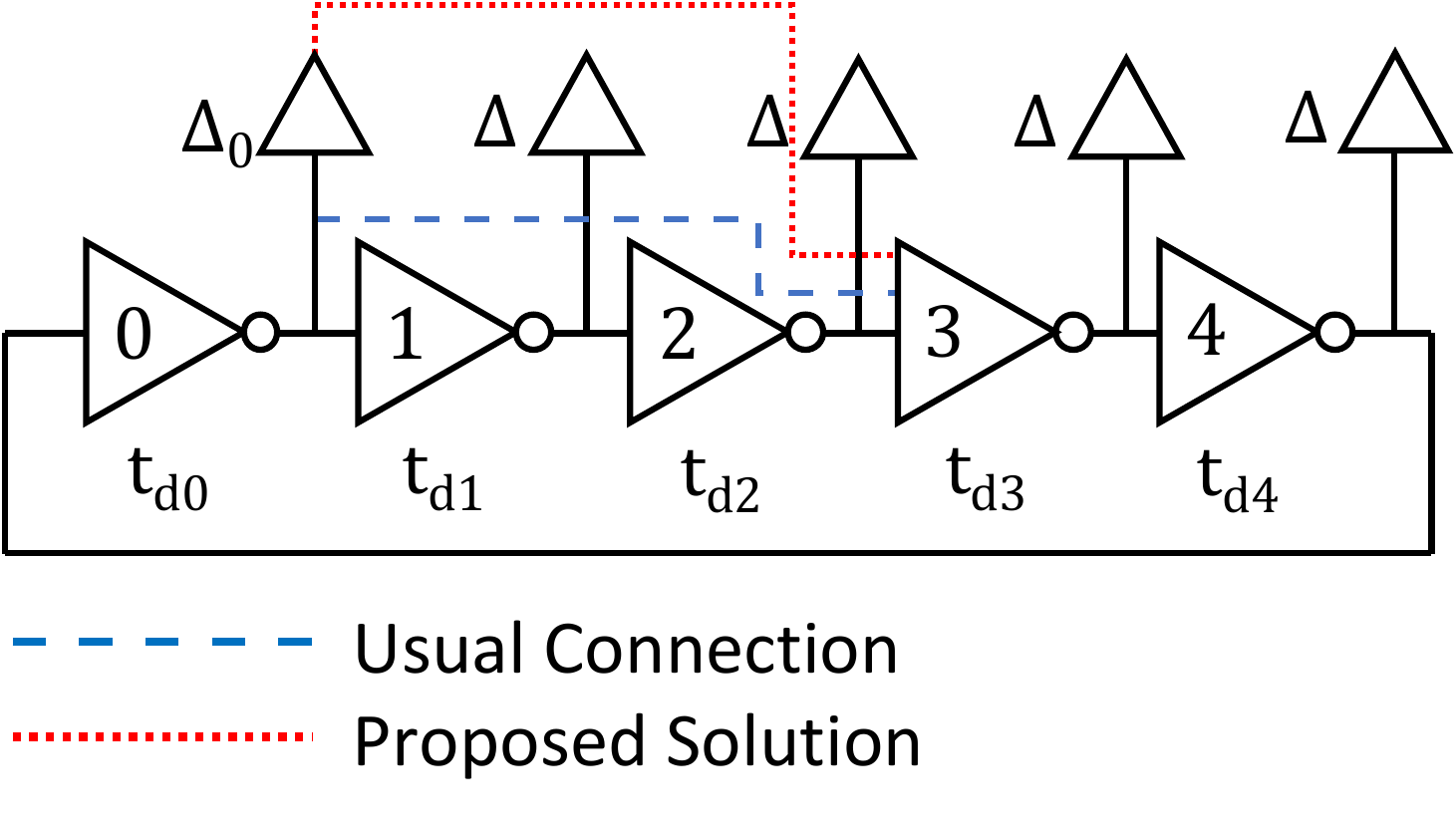}
\caption{An example of a 5 stage multi-path RO with two types of feedforward connections (dotted line: proposed solution). $\Delta_0\neq \Delta$ indicates the propagation time of the buffer that shows a mismatch with respect to the others.}
\label{example_dnl_feedforward}
\end{figure}

\begin{figure}[!t]
\centering
\includegraphics[width=3.3in]{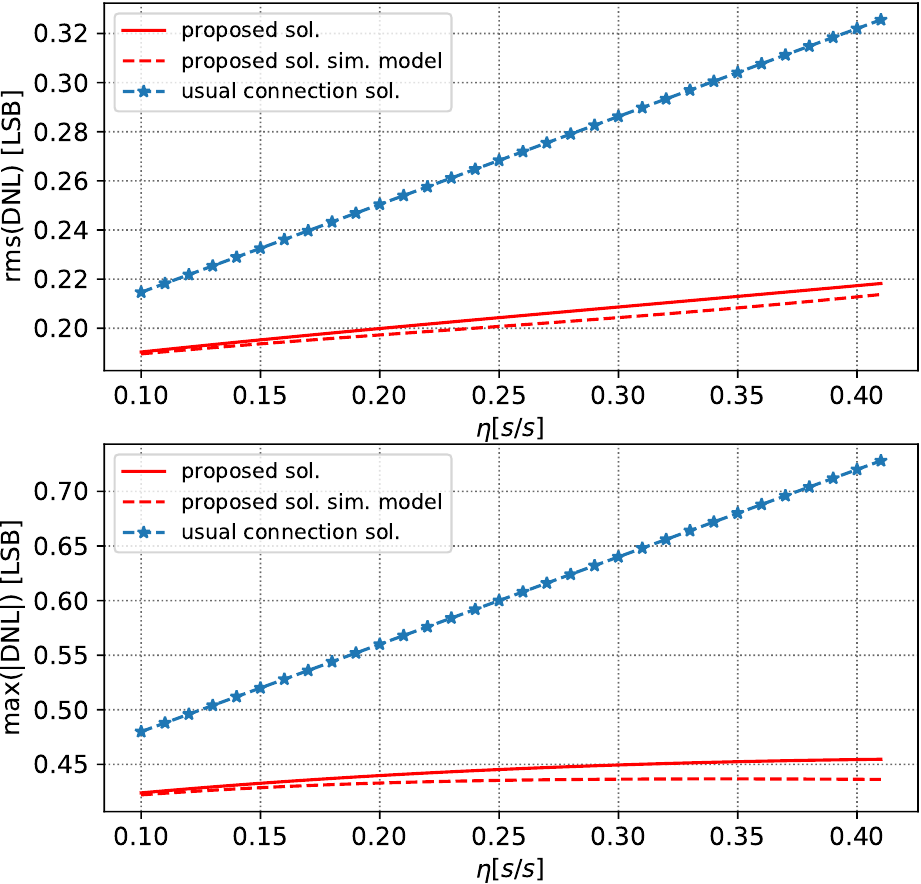}
\caption{RMS (top) and maximum of the absolute value (bottom) of DNL as function of $\eta$ of both of the solutions depicted in Figure \ref{example_dnl_feedforward} (calculated with Eq. \ref{2_eq:dnl_blue_case} for the usual connection case, with Eq. \ref{2_eq:dnl_red_case} for the proposed solution scenario and exploiting the edge time distribution of Eq. \ref{eq:detailed} for the more detailed model).}
\label{dnl_analysis}
\end{figure}

\begin{figure}[!t]
\centering
\includegraphics[width=3.3in]{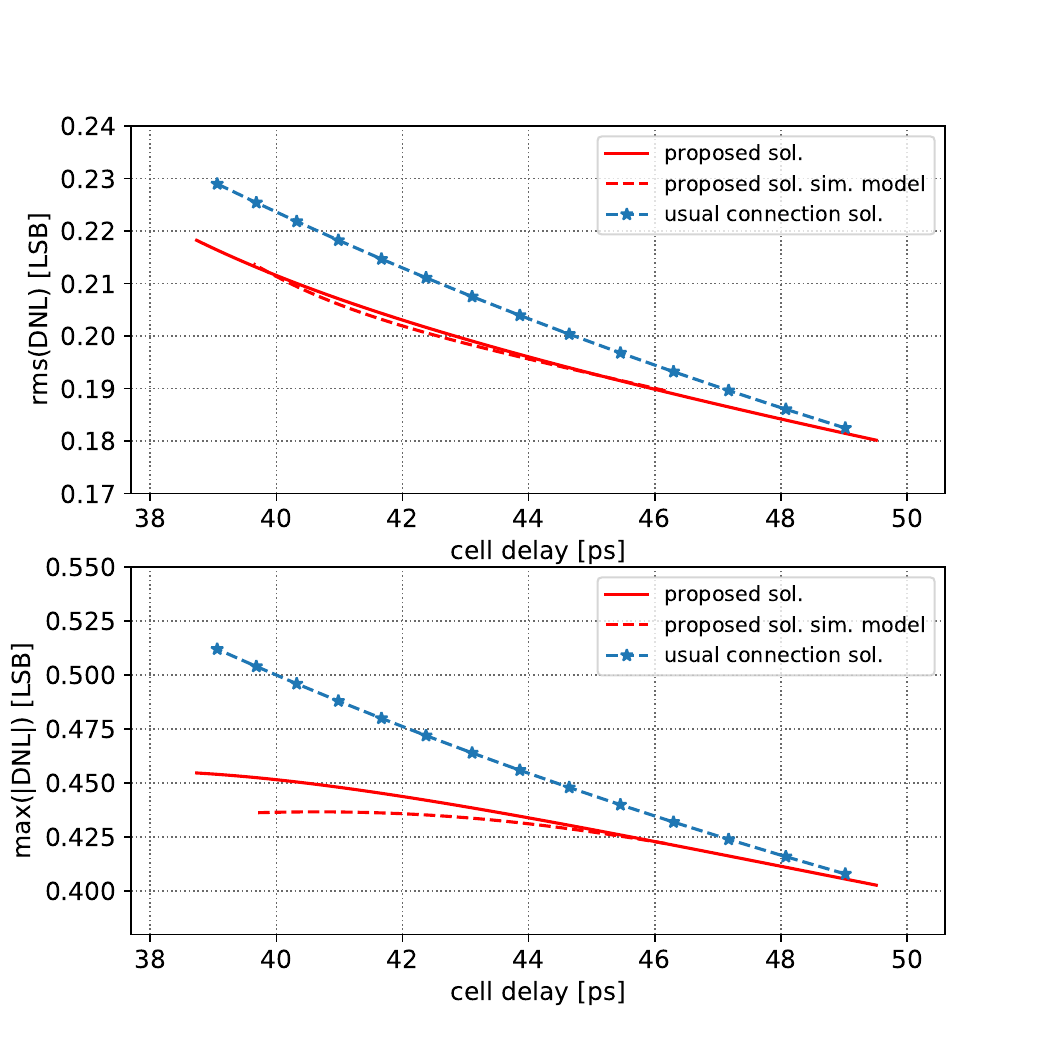}
\caption{RMS (top) and maximum of the absolute value (bottom) of DNL as function of the cell delay (calculated with Eq. \ref{2_eq:dnl_blue_case} for the usual connection case, with Eq. \ref{2_eq:dnl_red_case} for the proposed solution scenario and exploiting the edge time distribution of Eq. \ref{eq:detailed} for the more detailed model).}
\label{dnl_analysis_td}
\end{figure}

The presented TDC is composed of a RO with 9 pseudo-differential pseudo-NMOS delay cells, depicted in Figure \ref{delay_cell_buffer_first_case}. Each of the output pairs of these cells is connected to a pseudo-NMOS Differential Cascode Voltage-Switch-Logic (DCVSL) buffer \cite{turker2011dcvsl}, shown in Figure \ref{delay_cell_buffer_second_case}. The pseudo-NMOS architecture was chosen to increase the oscillator frequency: in this way, the load connected to each cell does not include the gate capacitances of PMOS transistors. \\In a conventional RO, the frequency of the output is given by the inverse of the time that signal needs to propagate inside the chain of delay cells multiplied by two:
\begin{equation}
    \label{2_eq:fro}
    f_{RO}=\frac{1}{2Nt_{d}},
\end{equation}
where $N$ is the number of stages of the oscillator and $t_{d}$ is the delay of the single stage that represents the limit in time-resolution of a TDC with a conventional RO. However a feedforward design (also indicated as multi-path) has been applied to increase the speed of the system, reducing the delay $t_d$ and, therefore, improving the resolution (LSB is given by $t_d$ as explained in Section \ref{sec:introduction}). Indeed, each delay cell of Figure \ref{delay_cell_buffer_first_case} features two differential inputs: one of them is connected to the output of the previous cell while the other to the outputs of the buffer related to the cell placed four stage before in the RO. In this way, each buffer will be used to advance the charge or the discharge of the input of a further cell, resulting in a consequent increase of the oscillation frequency, as shown in Figure \ref{ring_oscillator}. For this reason, as simulations show, the nominal $f_{RO}$ will rise of almost 45 $\%$ with the respect to the case in which the multi-path architecture is not adopted.
Moreover, the inputs of one of the delay cells must be inverted as displayed in Figure \ref{ring_oscillator2} in order to make the circuit properly oscillate by having an odd number of inverting stages. Indeed, because of the way the stages are connected (Figure \ref{ring_oscillator}), each output propagates in the chain without being inverted as depicted in Figure \ref{ring_oscillator2}. For this reason, the connection in blue of Figure \ref{ring_oscillator2} is fundamental to satisfy the Barkhausen oscillation criterion \cite{ro_online,ramazani2014cmos,sedra1998microelectronic}. The choice of having a single inversion was made to facilitate a better symmetry of the layout.
\\The role of the buffers is to decouple the output nodes of the RO and the loads of the circuit, i.e. the latch stages used to sample the state of the oscillator. However, in our design, these blocks are also put in the feedforward paths in order to increase the linearity of the converter and reduce the effect of mismatch among the buffers by exploiting the feedback loops of the oscillator. In order to clarify this point, it is possible to analyze the simple 5 stage multi-path RO depicted in Figure \ref{example_dnl_feedforward} (the result of the following analysis is general and can also be applied to structures with a larger number of stages). The dashed line represents the conventional multi-path architecture in which the feedforward is provided directly by the outputs of the delay cells. In the proposed RO, buffers provide the input to later delay cells through the dotted connections of Figure \ref{example_dnl_feedforward}. The following analysis aims to evaluate the effect of the mismatch of an output buffer on the linearity of the architecture in both of the scenarios depicted in Figure \ref{example_dnl_feedforward}.
\\The parameters $t_{di}$ with $i=0,1,...,4$ are the delay of the inverters of the oscillator while the (non-inverting) buffers show a nominal delay given by $\Delta$. In order to analyze the linearity of the system, it is possible to exploit the Differential Non-Linearity (DNL) defined as 
\begin{equation}
    \label{2_eq:dnl_ideal}
    DNL(i)=\frac{t_{di}-t_d}{t_d},
\end{equation}
where $i$ is the code of the converter and $t_d$ is the ideal delay which, as stated before, corresponds to the ideal LSB.
Considering the first case (dashed line connection) with ideal delays $t_{di}=t_d$ $\forall i$ and assuming that, because of mismatches, the delay of the first buffer is $\Delta_0\neq \Delta$, the DNL will be
\begin{subnumcases} {\label{2_eq:dnl_blue_case} DNL(i)=}
\frac{t_d+(\Delta_0-\Delta)-t_d}{t_d}=\frac{\Delta_0-\Delta}{t_d} & $i = 0$\label{2_eq:dnl_blue_case_0}\\
0 & $i\neq0$
\end{subnumcases}
since the $\Delta_0$ will only affect the value of DNL related to the first cell. More in detail, the mismatch $\Delta_0 \neq \Delta$ may possibly generate a bubble in the output code (see Section \ref{sec:measurements}). In the proposed example, it is possible to evaluate the DNL associated to the RO using Eq. \ref{2_eq:dnl_blue_case} only by assuming that an efficient bubble correction algorithm has been implemented. The same assumption will be used for the rest of the section.
\\The characterization of the behavior of the RO requires the introduction of a parameter that links the effect of the feedforward connections with the speed of the system. The value of $t_d$ is function of the difference between the arrival times of the inputs of each cell $\delta$. Expanding $t_d=t_d(\delta)$ in a Taylor series and neglecting all the components after the linear one\footnote{The approximation of Eq. \ref{2_eq:taylor}, as it will be explained later in the section, is justified by simulations. However, the analysis reported in this paper is general and can be easily extended to situations in which the non-linear terms are not negligible.}, we obtain
\begin{equation}
    \label{2_eq:taylor}
    t_d(\delta)\approx t_d(0)+\frac{dt_d}{d\delta}(0)\delta.
\end{equation}
From Figure \ref{example_dnl_feedforward}, it is possible to see that in the dashed line case $\delta=-2t_d$. Replacing this relation in Eq. \ref{2_eq:taylor} leads to  
\begin{equation}
    \label{2_eq:td_taylor}
    t_d=t_{dmax}-2\eta t_d \longrightarrow t_d=\frac{t_{dmax}}{1+2\eta},
\end{equation}
where $t_{dmax}=t_d(0)$ is the maximum value of $t_d$ (in the case of no multi-path architecture implemented) and $\eta=dt_d(0)/d\delta$ is the feedforward parameter described before. Simulations of the cell in Figure \ref{delay_cell_buffer_first_case} justify the approximations of Eq. \ref{2_eq:taylor} and \ref{2_eq:td_taylor} with values of $\eta \approx 0.25$. The star-marked curves of Figure \ref{dnl_analysis} show the behavior of the maximum and the Root Mean Square (RMS) value of the DNL as function of $\eta$ with $t_{dmax}=\Delta=50$ ps, $\Delta_0=70$ ps. For what concerns the proposed solution (dotted line in Figure \ref{example_dnl_feedforward}), a proper evaluation of the non-linearities in the case $\Delta_0 \neq \Delta$ can be performed analysing the distribution of the edge times in each node of the oscillator $t_i$. As done for Eq. \ref{2_eq:taylor} and \ref{2_eq:td_taylor} and considering the presence of the delay buffers in the feedforward paths, these times can be expressed as 
\begin{equation}
    \label{eq:detailed}
    t_{i+1}=t_{i}+t_{dmax}-\eta[t_i-(t_{(i-2)\bmod5}+\Delta_{(i-2)\bmod5})].
\end{equation}
A numerical approach was used to calculate the values of $t_i$ for enough oscillator cycles such that all delay cells $t'_{di}$ reach their convergence values. At this point, the DNL can be calculated exploiting Eq. \ref{2_eq:dnl_ideal}, replacing $t_d$ with the average value of the cell delays $t'_{d-av}$ and taking into account that $\Delta_0 \neq \Delta$ as done for Eq. \ref{2_eq:dnl_blue_case_0}.
The plots in Figure \ref{dnl_analysis} show that, for the proposed solution (dashed line curves), the RMS and the maximum of the absolute value of the DNL is smaller than the one related to the usual feedforward architecture (star-marked curves). The same parameters can also be compared as function of the cell delays (LSB). In Figure \ref{dnl_analysis_td}, it is possible to see that the non-linearity of the proposed solution has smaller values also when $t_d$ and $t'_{d-av}$ are comparable. The use of $t'_{d-av}$ instead of $t_d$ will be justified in Subsection \ref{sub:event_by_event}. Indeed, the TDC is featuring an event-by-event calibration system that is able to compensate potential variations in the oscillation period measuring the frequency of the RO through a comparison with an external reference signal.
\\A simplified approach can be used to analyze the behavior of the proposed solution. This approach is based on neglecting the variation of $t'_{di}$ as function of the variation of other cell delays and considering for it only the impact of $\Delta$. This simplification, as it will be later shown, will give similar results to the ones obtained with the more detailed approach explained before because, in this analysis, only the effect of the mismatch of the buffers has been evaluated. Following the same considerations that lead to Eq. \ref{2_eq:td_taylor}, it is possible to obtain the value of the cell delays $t'_d$ as
\begin{equation}
    t'_d=t_{dmax}-\eta(2t'_d-\Delta) \longrightarrow t'_d=\frac{t_{dmax}+\eta \Delta}{1+2\eta}. 
\end{equation}
However, the mismatch on the first buffer will also have an impact on the delay $t'_{d3} \neq t'_d$ that can be expressed as 
\begin{equation}
    \label{2_eq:td3}
    t_{d3}=t_{dmax}-\eta(2t'_d-\Delta_0)=t'_{d}+\eta(\Delta_0-\Delta).  
\end{equation}
The new value of $t_{d3}$ will also cause a variation in the oscillation period of the RO 
\begin{equation}
    \label{2_eq:Tro}
    T_{RO}=2[5t'_d+\eta(\Delta_0-\Delta)]. 
\end{equation}
From Eq. \ref{2_eq:Tro}, it is possible to obtain the value of the equivalent LSB of the system (i.e. the average elementary delay of the cells) as 
\begin{equation}
    t'_{d-av}=\frac{T_{RO}}{10}=t'_d+\frac{\eta}{5}(\Delta_0-\Delta).
\end{equation}
Thus, the DNL of the architecture will be given by 
\begin{subnumcases} {\label{2_eq:dnl_red_case} DNL(i)=}
\frac{(\Delta_0-\Delta)(1-\frac{\eta}{5})}{t'_{d-av}} & $i = 0$\label{2_eq:dnl_red_case_0}\\
\frac{-\frac{\eta}{5}(\Delta_0-\Delta)}{t'_{d-av}} & $i=1,2,4$\\
\frac{\frac{4}{5}\eta(\Delta_0-\Delta)}{t'_{d-av}} & $i=3$.
\end{subnumcases}
It must be clarified that in a $N$ stages RO-based TDC, the total number of different codes the system is able to provide as output is $2N$. Hence, the $DNL(i)$ should be defined for $i=0,1,...,2N-1$. 
However, in this simplified analysis, assuming that the rise and fall times of the cells are perfectly equal, the mismatches affect the value of $DNL(i)$ for $i=j$ and $i=j+N$ with $j=0,1,...,N-1$ in the same way. For this reason, it is possible to consider only half of the values of the DNL as done for Eq. \ref{2_eq:dnl_blue_case} and \ref{2_eq:dnl_red_case}.
\begin{figure}[!t]
\centering
\includegraphics[width=3.9in]{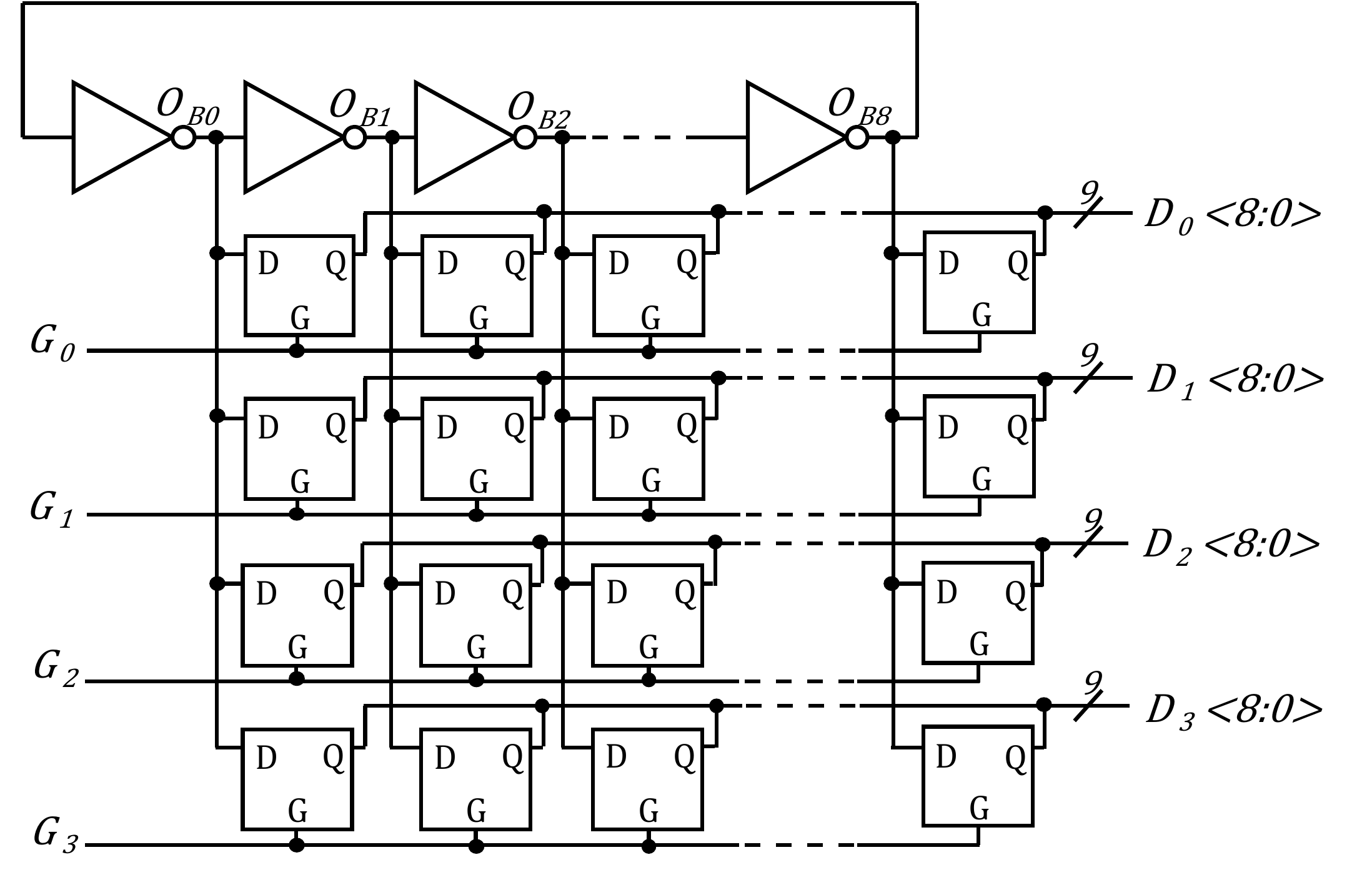}
\caption{Block diagram of the system for the event-by-event calibration.}
\label{ro_latch}
\end{figure}
In Figure \ref{dnl_analysis} and \ref{dnl_analysis_td}, the solid lines represent the behavior of the non-linearities of the architecture with this more simplified approach. The approximation of the previous analysis are negligible for low values of $\eta$ because of the reduced impact of the feedforward. However, even for larger $\eta$, the proposed solution shows better performance in terms of non-linearities. 
\\Finally, it must be emphasized that the choice of a differential architecture, despite the increase of power consumption, is also based on improving the linearity of the system: simulations show that the DNL of a single-ended solution is almost 14 $\%$ higher than the one of an equivalent differential structure. 
\begin{figure}[!t]
\centering
\includegraphics[width=2.5in]{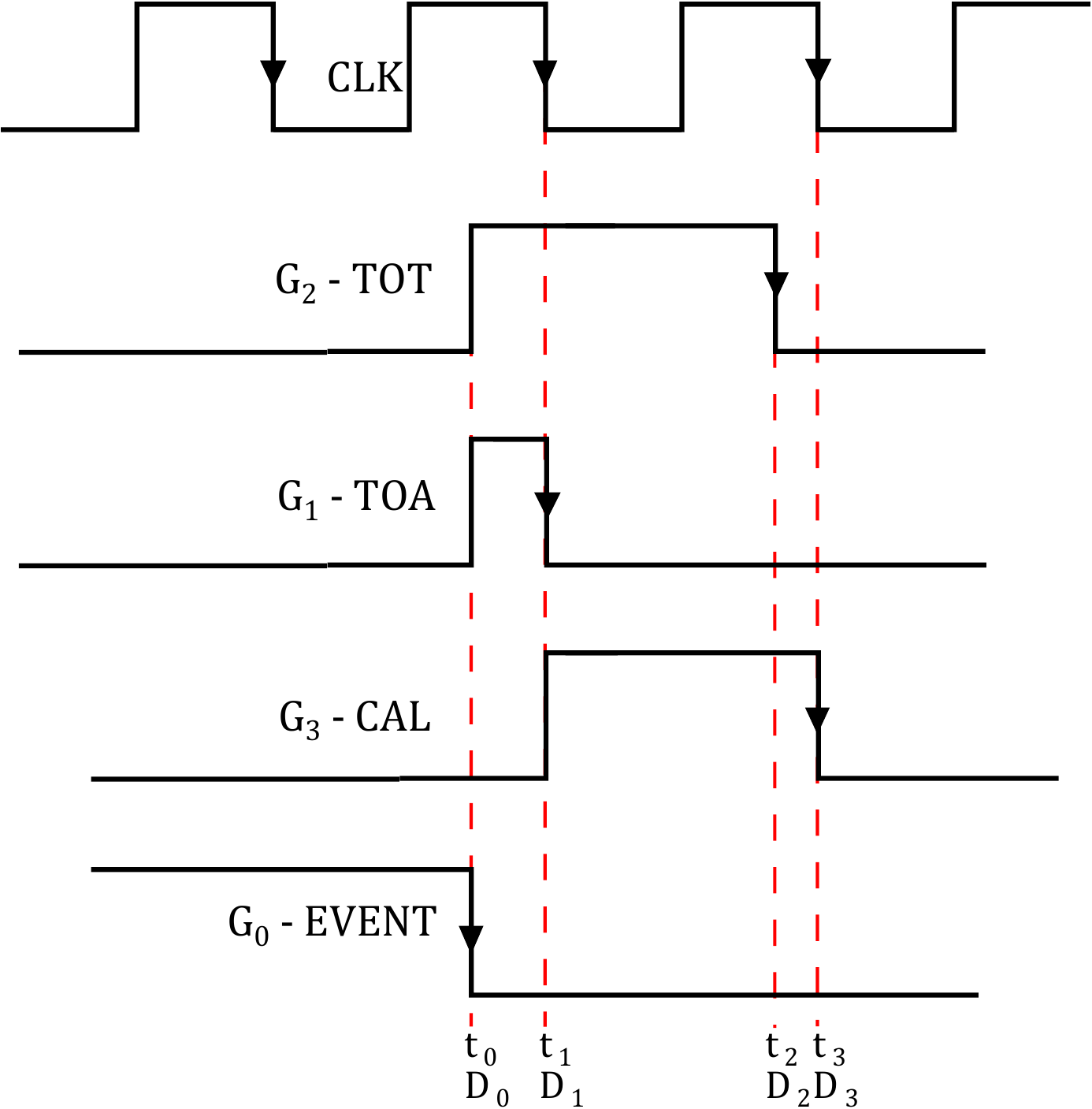}
\caption{Reference clock signal $CLK$ (up) and gating signals $G_j$ (down).}
\label{event_by_event_signals}
\end{figure}

\subsection{Event-by-event Measurement System}
\label{sub:event_by_event}
In Figure \ref{ro_latch} we describe the synchronization system to which the TDC is connected. This system is based on the one presented in \cite{cardarelli2018european}.
\\Each node of the RO $O_{Bi}$ with $i=0,1,...,8$ is connected to 4 stages of D-latch. Their outputs $D_j<8:0>$ with $j=0,...,3$ will follow the signals produced by the RO when the latches are in transparent mode (in this case when gating signals $G_j=1$). The falling edge of $G_j$ will lead latches in hold mode and sample the oscillator signals into the $D_j$ outputs. Three counters must be connected to as many outputs of the four latch stages. The gating signal $G_0$ is connected to the EVENT line, that will perform a falling edge every time an event occurs. A logic will then generate the remaining gating signals $G_{1,2,3}$ that, for image reconstruction applications, can be associated to Time-of-Arrival (ToA), Time-Over-Threshold (TOT) and the period of a reference clock (CAL) respectively (it must be highlighted that a different number of latch stages can be adopted for different types of applications in which the TDC can be used). The counters will calculate the number of oscillator cycles $N_{C}$ in these time intervals distributed as in Figure \ref{event_by_event_signals}, producing coarse measurements of these periods $T_{coarse}=N_{C}T_{RO}$. The difference between the states of the TDC at the beginning and at the end of ToA, TOT and CAL intervals will define the fine contributes of the measurements $T_{fine}=(D_i-D_j)t_d$ where $D_i$ and $D_j$ are the outputs of two of the latch stages and $t_d$ is the resolution of TDC (as stated before, it corresponds to the delay of the cells of the RO). From Figure \ref{event_by_event_signals}, considering both of the fine and coarse contributes and resolving the RO period as $T_{RO}=2Nt_d$ (with $N=9$ in this case), it is possible to express the ToA, TOT and CAL intervals as
\begin{gather}
    T_{ToA}=t_d[N_{C1}2N+(D_1-D_0)] \label{2_eq:measurements_event_toa}\\
    T_{TOT}=t_d[N_{C2}2N+(D_2-D_0)] \label{2_eq:measurements_event_tot}\\
    T_{CAL}=t_d[N_{C3}2N+(D_3-D_1)] \label{2_eq:measurements_event_cal}
\end{gather}
The measurement of $T_{CAL}$ is fundamental to compensate for potential parasitics, device mismatches, voltage drops of the supply, temperature gradients and in general all those factors that may cause a variation of the $t_d$ and a consequent worsening of the accuracy of the converter.
\begin{figure}[!t]
\centering
\includegraphics[width=2.5in]{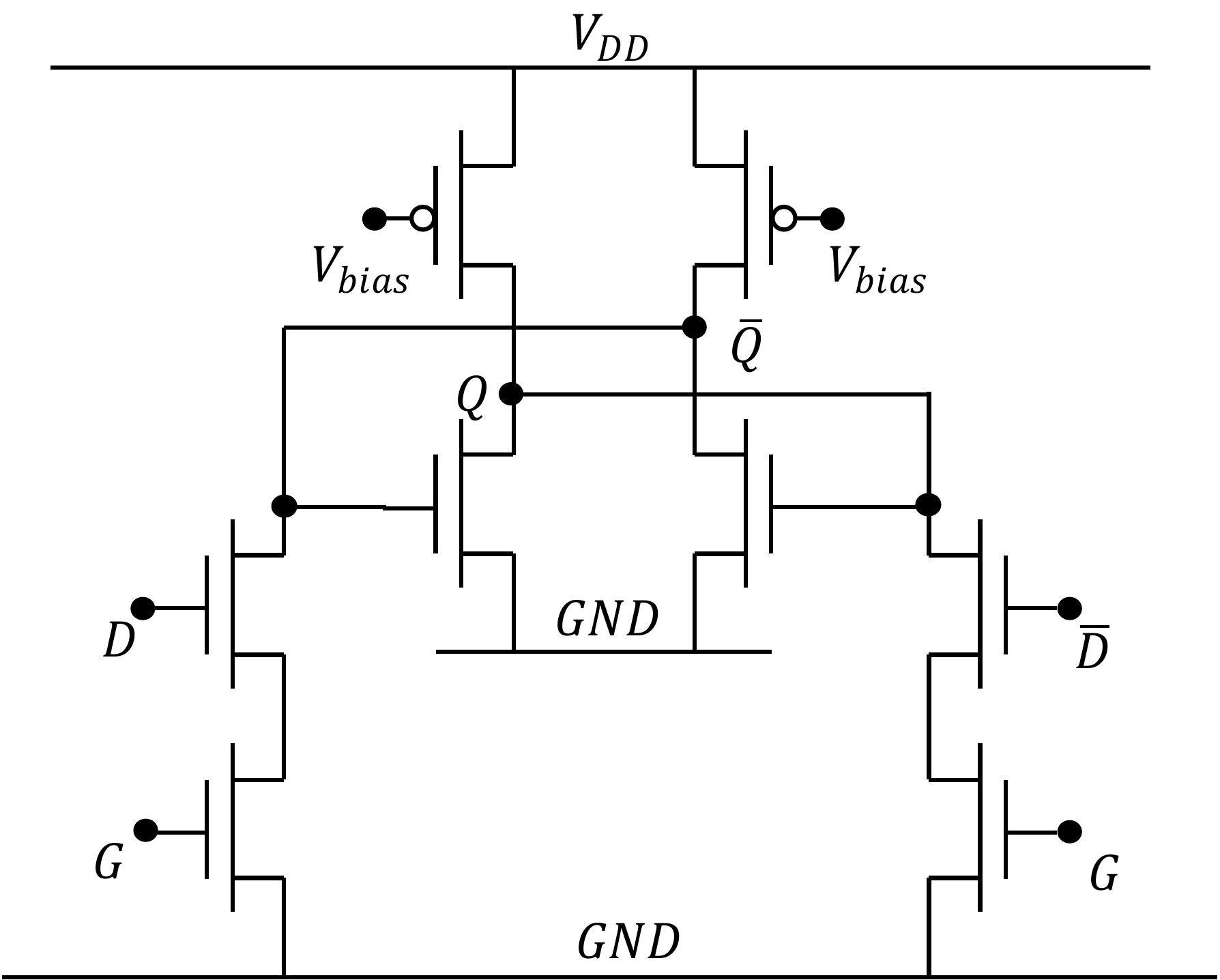}
\caption{Schematic of the latches used to sample the state of the RO.}
\label{latches}
\end{figure}
\begin{figure}[!t]
\centering
\subfloat[]{\includegraphics[width=2in]{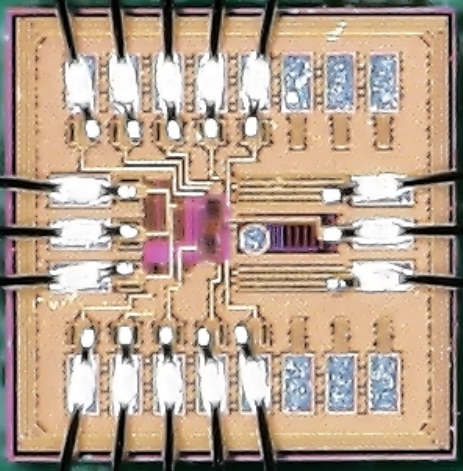}%
\label{layout_chip}}
\hfil
\subfloat[]{\includegraphics[width=2in]{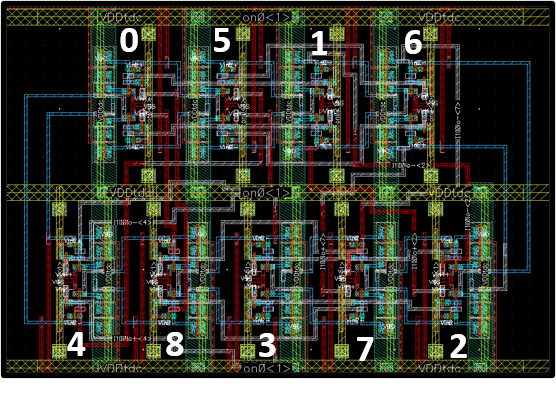}%
\label{layout_layout}}
\caption{Picture of the test chip of the proposed TDC (total area: 0.9 x 0.9 mm\textsuperscript{2}) (a) and layout of the RO (b).}
\label{layout}
\end{figure}
Indeed, the value of $T_{CAL}$ is nominally equal to an external clock reference. For this reason, Eq. \ref{2_eq:measurements_event_cal} can be exploited to calculate the value of $t_d$ as function of the clock period every time an event occurs. Hence, this approach allows avoiding the use of any PLL-based synchronization system reducing the complexity of the whole architecture, power consumption and noise. The value of the LSB, i.e. $t_d$, can vary in time due to the above-anticipated temperature effects. This system, however, is able to calculate this value in a time window that depends on the period of the reference signal ($T_{CAL}$), allowing the TDC to provide an output coherent with the time to be measured. Moreover, in a chip with many ROs and only one PLL, all the frequencies would be synchronized on the slowest one. The approach shown above, instead, allows avoiding this situation, since all the ROs will oscillate at their own natural frequency. 
\\The schematic of the latches chosen for this architecture is depicted in Figure \ref{latches}. Also in this case, the pseudo-NMOS architecture has been chosen to reduce the propagation time of these blocks and make them able to follow the outputs of the RO ($D$ signals in Figure \ref{latches}) when the latches are in transparent mode.
\\A test chip of the TDC featuring one channel (i.e. 4 latch stages) was submitted and its measurements will be presented in Section \ref{sec:measurements}. A simulation analysis highlighted that the RO can be connected to more than one channel. Its oscillation frequency is reduced by 5.5\% if 2 channels are connected and 23\% in the case of a 4 channels configuration. However, in the applications in which such a drop is not acceptable, it is possible to add more ROs and/or multiplex more pixels to the same TDC channel. The integration of multiple ROs is usually problematic for area and power consumption. However, as it will be shown in Section \ref{sec:layout}, \ref{sec:measurements_subsec} and \ref{sec:state-of-the-art-comparison}, the area and the dissipated power of the proposed architecture is smaller or comparable to the ones of many state-of-the-art TDCs.
\\The jitter of the CLK signal of Figure \ref{event_by_event_signals} directly affects the precision of the measurement. In the proposed solution, since a $\approx$30 ps LSB is achieved, a jitter in the order of few ps is required. The distribution of a clock with a picosecond level jitter in a large ASIC is a challenging task in terms of area and power consumption. Fortunately, a reference signal can be sent only when a calibration is necessary, so the clock can be gated for a majority of the time, sending it only when an event is detected or at a fixed rate, depending on the expected drift in frequency of the clock source. 
\begin{figure}[!t]
\centering
\includegraphics[width=4in]{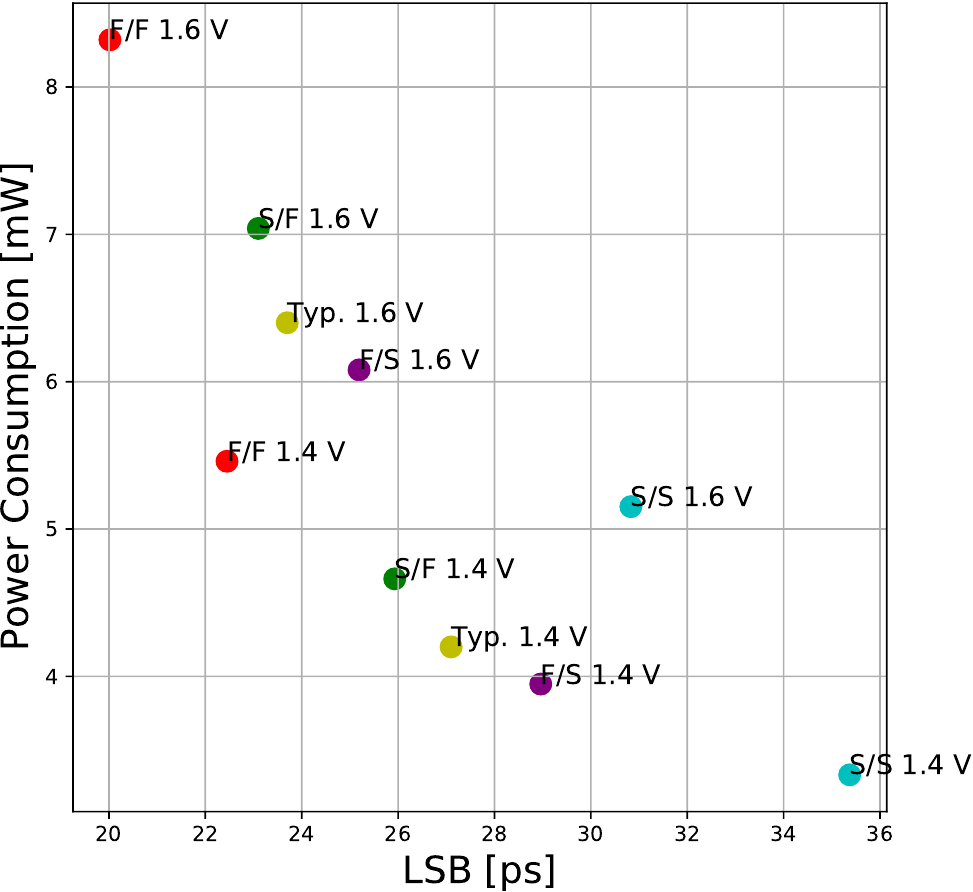}
\caption{LSB and power consumption of the TDC for typical, Fast/Fast (F/F), Fast/Slow (F/S), Slow/Fast (S/F) and Slow/Slow (S/S) corners and for $V_{DD}$ equal to 1.4 V and 1.6 V.}
\label{corner_tdc_mos}
\end{figure}

\subsection{Layout}
\label{sec:layout}
A picture of a test chip for the proposed TDC is shown in Figure \ref{layout_chip}, while Figure \ref{layout_layout} shows the layout of the RO. The position of the delay cells and buffer has been chosen to maximize the symmetry of the connections. As it is possible to see in the figure, with this placement the lengths of the feedforward paths are always one cell long while direct paths are two. The area of the RO core is 30.1 $\upmu$m x 20.9 $\upmu$m and 30.1 $\upmu$m x 87.5 $\upmu$m including the rest of the the system. Moreover, the outputs of the latches connected to the RO are routed on different metal layers (the pattern is 5-1-3-1-3-5 for the three inner stages) in order to reduce capacitive couplings and their effect on oscillation frequency.

\begin{figure}[!b]
\centering
\includegraphics[width=4in]{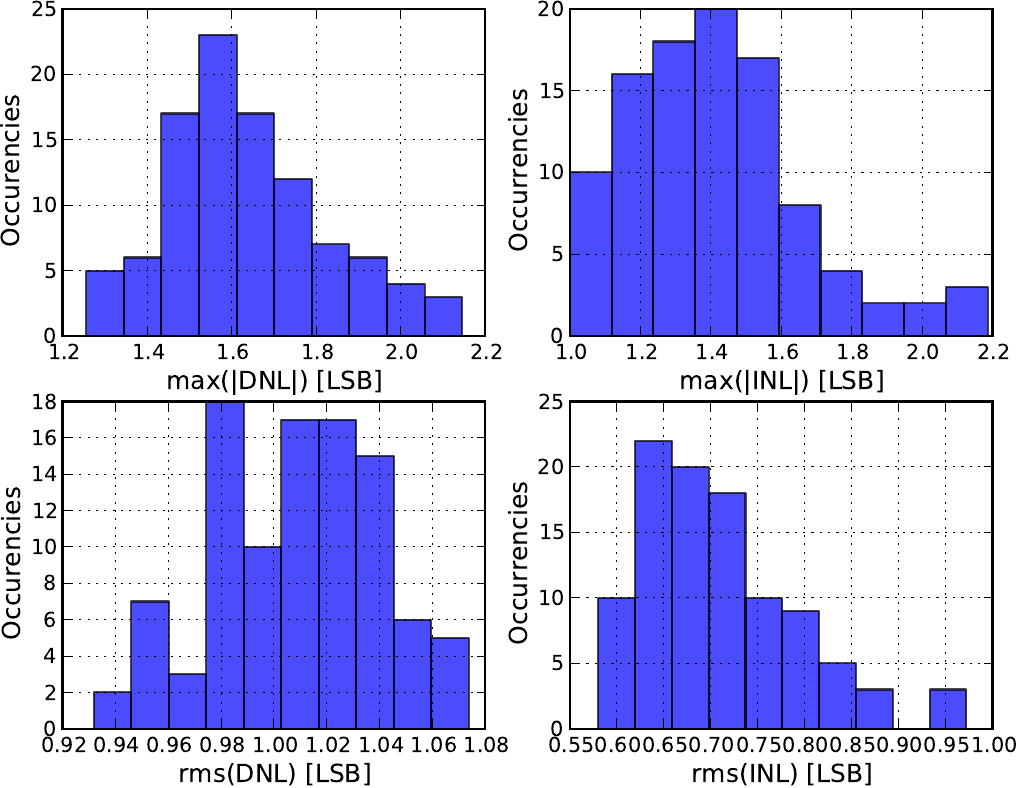}
\caption{Maximum values and RMS distributions of DNL and INL calculated over various Monte Carlo simulations. In this case, the supply $V_{DD}=1.6$ V.}
\label{dnl_inl_1_6}
\end{figure}

\section{Simulations and Measurements}
\label{sec:measurements}

In this section the simulations and the measurements of a test chip of the TDC will be shown. As stated before, the converter was designed in 130 nm CMOS technology and the simulation framework was set to analyze and optimize the performance of the circuit in terms of scalability, linearity and time-resolution.

\subsection{Post-layout Simulations}
\begin{figure}[!b]
\centering
\includegraphics[width=4.2in]{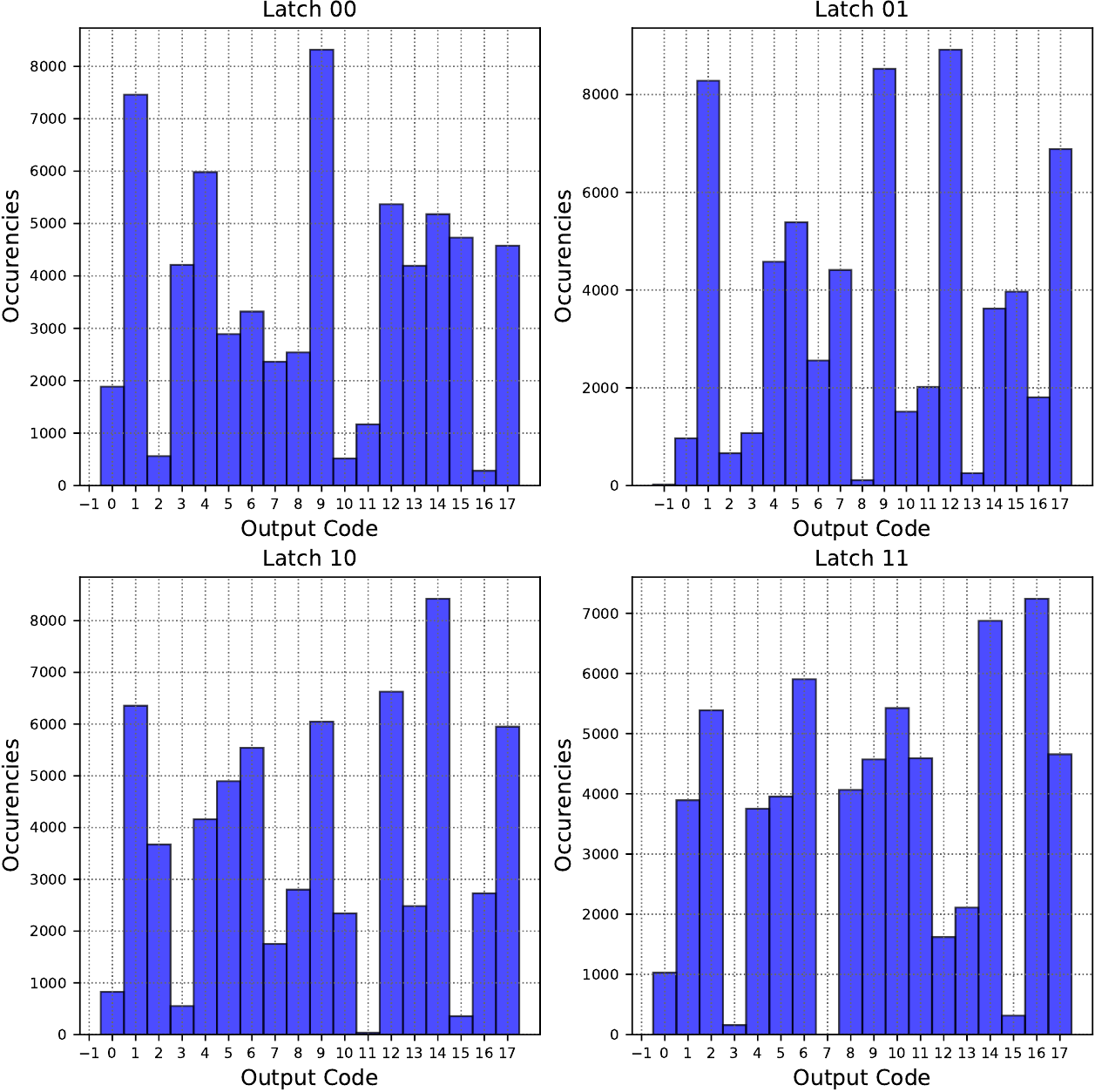}
\caption{Measured output distribution (after correction) of the TDC for $V_{DD}=1.6$ V and for all the latch stages connected to the RO.}
\label{dnl_measurements_1_6}
\end{figure}
The free-running frequency of the oscillator $f_{RO}$ is highly dependent on the parasitics of the system. Simulations highlighted a 61 $\%$ drop (on average) of the $f_{RO}$ when passing from schematic to post-layout netlist. The circuit has been analyzed for various supply voltages $V_{DD}$ with a focus on 1.4 V and 1.6 V. Post-layout simulations show that the RO oscillates at a frequency $f_{RO}$ equal to 2.05 GHz and 2.34 GHz for $V_{DD}=1.4$ V and $V_{DD}=1.6$ V respectively. Considering Eq. \ref{2_eq:fro} with $N=9$, the system will be characterized by a nominal resolution of 27.1 ps and 23.7 ps for the above-mentioned cases. Multi-corner simulations highlighted a less than 30 $\%$ variation of the LSB with the respect to the typical case. More in detail, minimum values of the LSB are obtained in Fast/Fast corner (22.45 ps and 20.02 ps for $V_{DD}=1.4$ V and $V_{DD}=1.6$ V respectively) and the maxima in the Slow/Slow (30.38 ps and 35.37 ps for $V_{DD}=1.4$ V and $V_{DD}=1.6$ V respectively). These values are reported in the plot of Figure \ref{corner_tdc_mos}.
\\A preliminary analysis has been performed during the design process to evaluate the linearity of the system. The sampling of the RO was simulated sweeping the sampling time $t_s$ in a time interval that is larger than $T_{RO}$, in order to be sure that the the system goes through all of its $2N$ states. The time step for $t_s$ was chosen equal to 1 ps. For each step, several Monte Carlo (MC) simulations have been performed (using the same set of seeds for every value of $t_s$, in order to make the outputs coherent). At this point, it is possible to calculate the DNL and the Integral Non-Linearity (INL) in order to evaluate the distribution of their maximum values and RMS. The INL can be defined as the integral of the DNL
\begin{equation}
    INL(i)=\sum_{n=0}^{i} DNL(n). 
\end{equation}
The distribution of the DNL and INL obtained through this analysis for the case $V_{DD}=1.6$ V is reported in Figure \ref{dnl_inl_1_6}. Table \ref{tdc_tab:post_layout_sim} shows the value of frequency, nominal resolution, power consumption and average value of both DNL and INL distribution (maximum value and RMS). The table also reports the simulated conversion time $T_{conv}$. This parameter (equal to approximately 0.69 ns and 0.51 ns for $V_{DD}=$1.4 V and 1.6 V respectively) only takes into account the time needed by the system to sample the state of the RO and the delay of the registers of the counters included in the converter. Thus, it represents the minimum ideal conversion time of the system. The measurement setup of the TDC, that will be described in the next subsection, did not allow a correct estimation of the conversion time since the system was limited by the readout logic. Hence, the aforementioned values of Table \ref{tdc_tab:post_layout_sim} just give an indication of the potential speed of the proposed TDC. Moreover, the $T_{conv}$ of the converters presented in the cited works (whose performance will be later commented and compared to our work) were extracted from the output data rate of the TDCs reported on the papers. Therefore, they simply represent upper limits of the real conversion times.

\afterpage{%
\clearpage
\begin{landscape}
\begin{table}
\renewcommand{\arraystretch}{1.3}
\caption{Multi-path simulations and measurements results. A comparison with other works is also reported.}
\label{tdc_tab:post_layout_sim}
\centering
 \begin{threeparttable}
        \begin{tabular}{ccccc|p{1cm}p{1cm}p{1cm}p{2cm}p{1cm}p{1cm}p{1cm}}
        \hline 
         & \multicolumn{2}{c}{Sim.} & \multicolumn{2}{c|}{Meas.} & \cite{swann2004100} & \cite{nissinen2003cmos} & \cite{kim2018two} & \cite{muntean2020blumino} & \cite{andersson2014vernier} & \cite{park2011cyclic} & \cite{liscidini2009time}\tabularnewline
        \hline 
        Architecture & \multicolumn{4}{c|}{PLL-less Multi-path RO} & TAC\tnote{1} & RO & RO-TA\tnote{2} & Multi-path RO & Vernier line & Cyclic Vernier & 2-D Vernier\tabularnewline
        \hline 
        V\textsubscript{DD} {[}V{]} & 1.4 & 1.6 & 1.4 & 1.6 & 5 & 3 & 1.8 & 3.3 & 1.2 & 1 & 1.2\tabularnewline
        Technology [nm] & \multicolumn{4}{c|}{130} & 500 & 350 & 180 & 180 & 65 & 65 & 65\tabularnewline
        Area {[}mm\textsuperscript{2}{]} & \multicolumn{4}{c|}{0.0006 (0.0026)\tnote{3}} & 2.88 & 3.27 & 0.34 & - & 0.0036 & 0.0064 & 0.02\tabularnewline
        LSB {[}ps{]} & 27.1 & 23.7 & 38.7 & 33.6 & 312 & 156 & 10.5 & 128 & 5.7 & 5.5 & 4.8\tabularnewline
        DNL\textsubscript{max} {[}LSB{]} & 1.41 & 1.63 & 1.34 & 1.26 & 0.2 & - & 0.7 & 5 & $<$1.5 & 1 & $<$1\tabularnewline
        INL\textsubscript{max} {[}LSB{]} & 1.28 & 1.41 & 1.77 & 2.02 & 0.3 & 0.23 & 0.5 & 2.4 & $<$9 & 1 & 3.3\tabularnewline
        DNL\textsubscript{RMS} {[}LSB{]} & 0.87 & 1.01 & 0.68 & 0.66 & - & - & - & - & - & - & -\tabularnewline
        INL\textsubscript{RMS} {[}LSB{]} & 0.57 & 0.66 & 1.15 & 1.31 & - & - & - & - & - & - & -\tabularnewline
        Power {[}mW{]} & 4.2 & 6.4 & 3.6 & 5.4 & 175 & 72 & 1.34 & 9 (1)\tnote{4} & 1.75 & 0.63 & 1.7\tabularnewline
        $\upsigma$\textsubscript{q} {[}ps{]}\tnote{5} & - & - & 21.1 & 17.1 & 100 & - & - & - & - & - & -\tabularnewline
        SSP {[}ps{]} & - & - & 15.8 & 19.5 & - & 78.5 & - & 57.6 - 98.6\tnote{6} & $<$17.1 & 2.31 & -\tabularnewline
        Accuracy {[}ps{]} & - & - & 40.9 & 31.0 & - & - & - & - & - & - & -\tabularnewline
        T\textsubscript{conv} {[}ns{]}\tnote{7} & $\approx$0.69 & $\approx$0.51 & - & - & 100 & - & 20\tnote{8} & - & 10\tnote{8} & - & 20\tnote{8}\tabularnewline
        E\textsubscript{conv} {[}mW$\cdot$ns{]} & $\approx$2.9 & $\approx$3.3 & - & - & 17500 & - & 26.8 & - & 17.5 & - & 34 \tabularnewline
        PN @ 100 kHz {[}dBc/Hz{]} & - & - & -97.7 & -99.6 & - & - & - & - & - & - & -\tabularnewline
        \hline 
        \end{tabular}
        \begin{tablenotes}\footnotesize
            \item[1] Time-to-Amplitude Converter.
            \item[2] RO Time Amplifier.
            \item[3] RO core (whole structure).
            \item[4] Peak (standby).
            \item[5] In \cite{swann2004100} indicated as resolution.
            \item[6] Min. and max. value reported on the paper.
            \item[7] For the proposed solution, it does not take into account the counters.
            \item[8] Extracted from the reported conversion rate.
        \end{tablenotes}
    \end{threeparttable}    
\end{table}
\end{landscape}
\clearpage
}

\subsection{Test Chip Measurements}
\label{sec:measurements_subsec}
The measurements of the test chip were performed using the UNIGE USB3 GPIO board, developed by the engineers of the Department of Nuclear Physics (DPNC) at University of Geneva and based on the architecture of the readout scheme of the Baby-MIND experiment detectors at CERN \cite{noah2016readout}. A firmware was loaded on the FPGA that the board features in order to handle the communication with the chip and send sampling signals for the analysis of the linearity of the TDC.

\subsubsection{Linearity Measurements and Bubble Correction}

\begin{figure}[!b]
\centering
\includegraphics[width=1.5in]{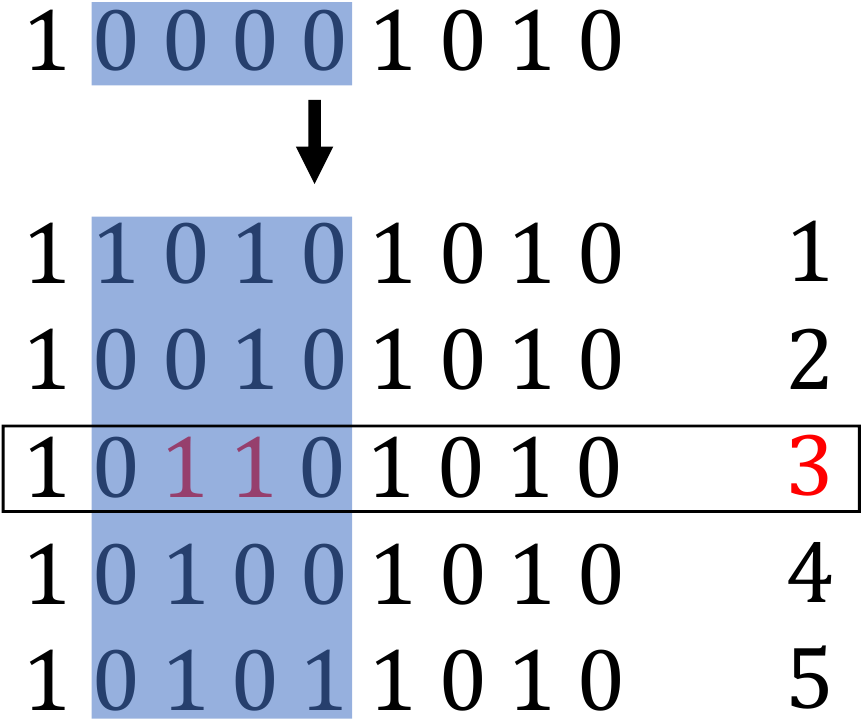}
\caption{Simple bubble correction algorithm implemented for the presented TDC. If four consecutive bits are 0 (word on top), assuming that the others are correct, there are only 5 possible states in which the RO can be (bottom). The numbers on the right represent the associated code (arbitrary) and they are ordered in the way the TDC goes through these states (e.g. 2 follows 1). The implemented correction is based on inverting the two middle bits of the incorrect portion of the word (in the full rectangle) because it reduces the maximum potential error and it is also the most probable value (proved after a simulation analysis).}
\label{bubble}
\end{figure}
\begin{figure}[!b]
\centering
\includegraphics[width=4in]{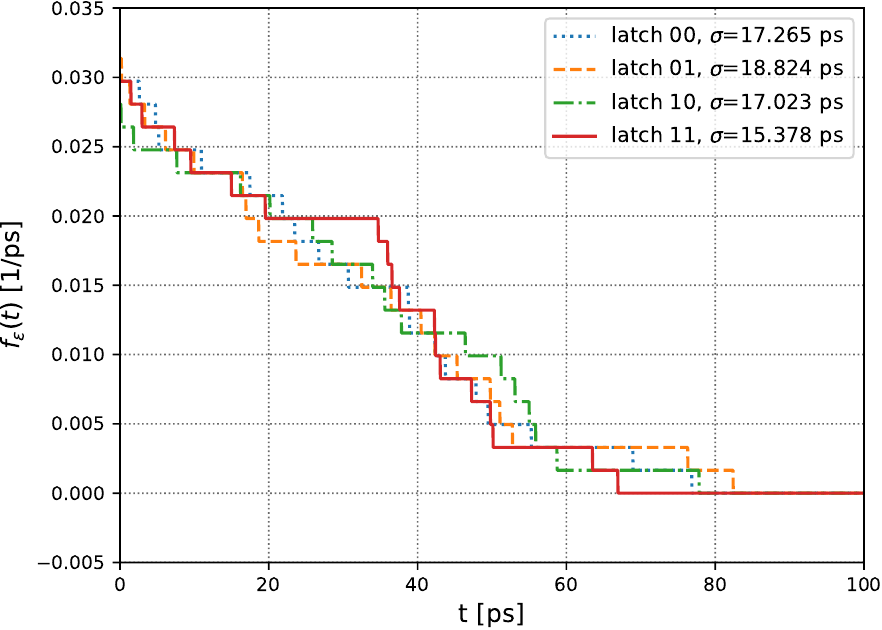}
\caption{Probability density function of the quantization error for each latch stage ($V_{DD}=1.6$ V).}
\label{pdf1_6}
\end{figure}

The distribution of the output read from all the latch stages connected to the RO after bubble correction is shown in Figure \ref{dnl_measurements_1_6} for $V_{DD}=1.6$ V. With bubble correction, it is possible to indicate the algorithms that can be exploited when a TDC provides a forbidden output. Indeed, a TDC as the one presented in this paper, features $N$-bit outputs but the number of correct states of the RO is only $2N$ \cite{henzler2010time}. However, because of mismatches and metastability of the latches, it is possible that the sampled word is not included among the $2N$ correct states and it is characterized by a group of more than two consecutive equal bits called bubble \cite{henzler2010time}. For the presented TDC, a simulation analysis highlighted that the most probable bubbles are the ones in which the output words has four consecutive zeros or ones and they can be easily corrected as explained in Figure \ref{bubble}. Applying this algorithm to the outputs obtained during the measurements it is possible to see that only the 0.03$\%$ of them is not corrected. In Figure \ref{dnl_measurements_1_6}, the output codes have been reported along the x-axis using numbers from 0 to 17 ($2N$) while -1 indicates the amount of forbidden state outputs after the correction (see the plot for latch 01).
\\Table \ref{tdc_tab:post_layout_sim} reports the results of the measurements, compared to the ones obtained with post-layout simulations. The test chip shows a smaller oscillation frequency that turns in to a lower time resolution due to non-extracted substrate capacitances that reduced the speed of the system. The measured LSB is 38.7 ps for $V_{DD}=1.4$ V and 33.6 ps for $V_{DD}=1.6$ V. However, the behavior of the circuit in terms of linearity is in line with the simulation results. 
\\The output distribution, as the one of Figure \ref{dnl_measurements_1_6}, allows calculating the standard deviation of the quantization error $\sigma_q$. This parameter can not be calculated using Eq. \ref{1_eq:sigma_ideal} because of the irregular and not ideal distribution of the bins of the system. The probability density function $f_\epsilon(t)$ of the error can be obtained using the law of total probability as 
\begin{equation}
    f_\epsilon(t)=\sum_{i=0}^{2N-1}f_\epsilon(t|C=i)P(C=i)
\end{equation}
where $P(C=i)=t_{di}/T_{RO}$ is the probability that the output code $C$ is equal to $i$. The behavior of the pdf for all the latch stages is reported in Figure \ref{pdf1_6} for $V_{DD}=1.6$ V. The average value of the quantization error standard deviation $\sigma_q$ is 21.1 ps (0.54 LSB) for $V_{DD}=1.4$ V and 17.1 ps (0.51 LSB) for $V_{DD}=1.6$ V. 

\begin{figure}[!b]
\centering
\includegraphics[width=4.5in]{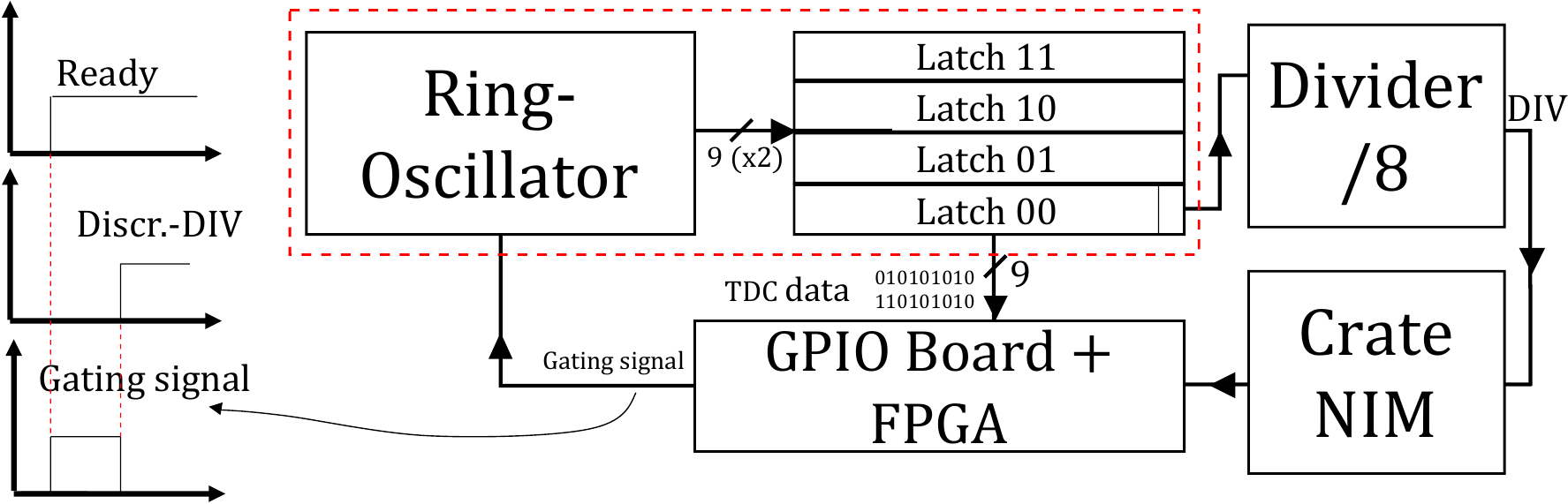}
\caption{Block diagram of the measurement system to evaluate the SSP of the converter.}
\label{diagram_ssp}
\end{figure}

\subsubsection{SSP and PN}

\begin{figure}[!t]
\centering
\includegraphics[width=4.5in]{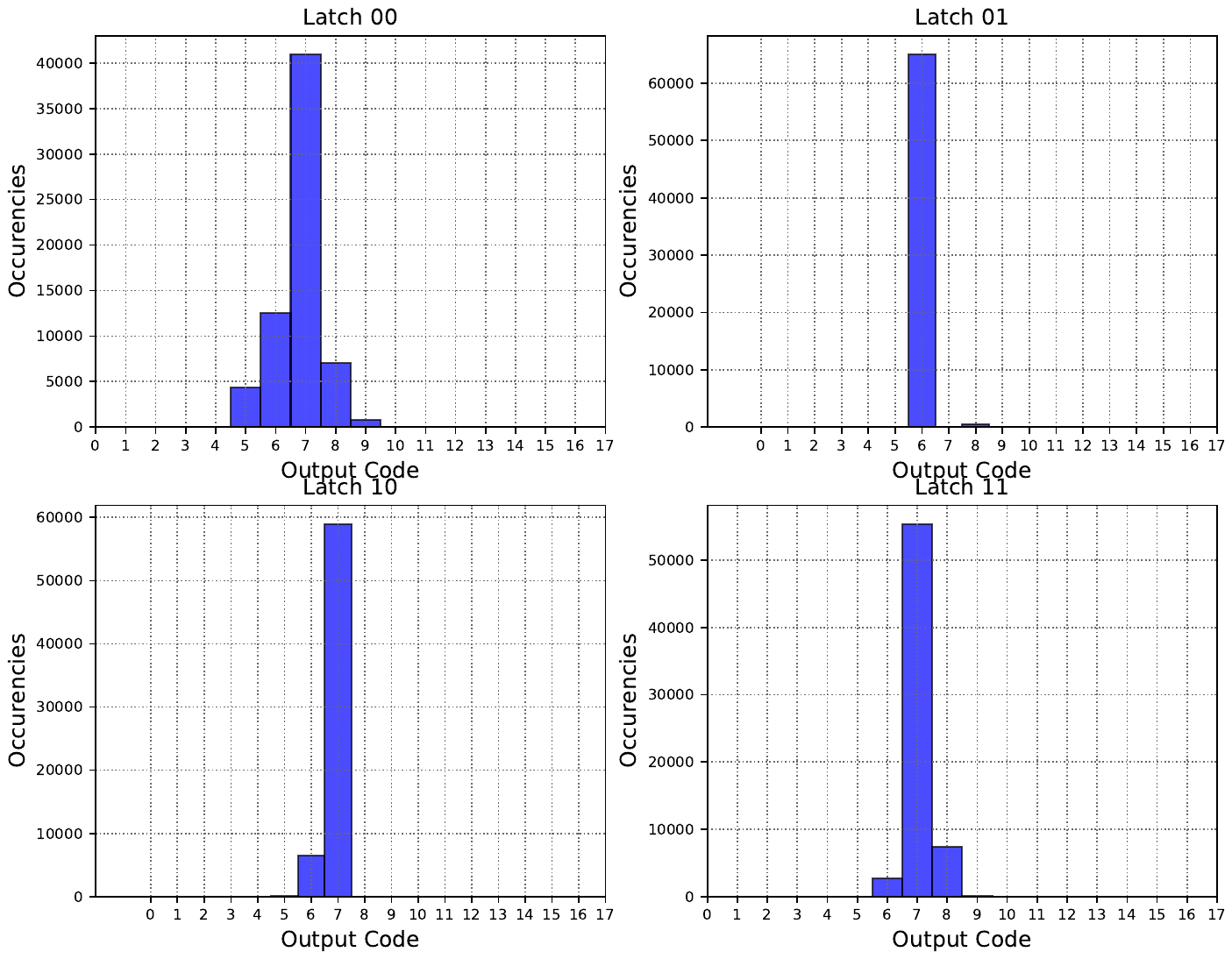}
\caption{Output distribution of the data obtained with the measurement system depicted in Figure \ref{diagram_ssp} for $V_{DD}=1.4$ V.}
\label{ssp_1_4}
\end{figure}

The so-called Single Shot Precision (SSP), i.e. the jitter of repeated measurements of the same time interval, was measured exploiting the block diagram in Figure \ref{diagram_ssp}. A Ready signal, connected to the gating of the latches, activates a 8 bit divider. The rising edge of the output of this block (Discriminated-DIV in the figure) is sent, through a Crate NIM, to the GPIO Board, that will then turn off the gating signals sampling the oscillator. The value provided by the TDC should ideally be always the same. However, the standard deviation of the distribution of this outputs will represent the above mentioned SSP. The output distribution for a supply voltage $V_{DD}=1.4$ V is reported in Figure \ref{ssp_1_4}. The average standard deviations are 15.8 ps (0.41 LSB) and 19.5 ps (0.58 LSB) for $V_{DD}=1.4$ V and $V_{DD}=1.6$ V respectively.
\begin{figure}[!t]
\centering
\includegraphics[width=4in]{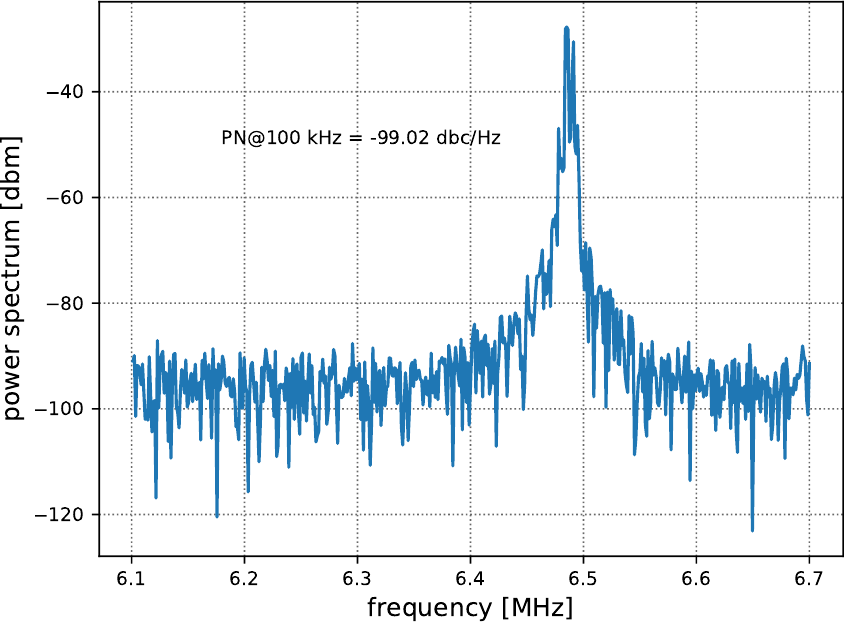}
\caption{Zoom of the power spectrum of the divider output for $V_{DD}=1.6$ V around the fundamental component of the signal.}
\label{PN}
\end{figure}
The analysis of the output distributions like the ones in Figure \ref{ssp_1_4} allows calculating the accuracy of the converter. This parameter can be defined as the equivalent offset affecting the time measuring system. For the presented TDC, the accuracy was evaluated as the maximum difference of the average value of the distributions obtained for the calculation of the SSP. The measurements show that the accuracy is equal to 40.9 ps (1.05 LSB) for $V_{DD}=$1.4 V and 31.0 ps (0.92 LSB) for $V_{DD}=$1.6 V. However, a simple calibration based on the same procedure implemented for the evaluation of the accuracy can be used for the offset compensation. 
\\The output of the divider was also exploited to analyze the power spectrum of the RO in order to evaluate the Phase Noise (PN). Figure \ref{PN} shows a zoom of the power spectrum of this signal around its fundamental component for $V_{DD}=1.6$ V. The measured value of PN at 100 kHz from this component is -99.02 dBc/Hz for a 1.6 V supply and -97.7 dBc/Hz for 1.4 V. The value of SSP and PN are reported in Table \ref{tdc_tab:post_layout_sim}.

\subsection{State-of-the-Art Comparison}
\label{sec:state-of-the-art-comparison}
\begin{figure}[!t]
\centering
\includegraphics[width=4in]{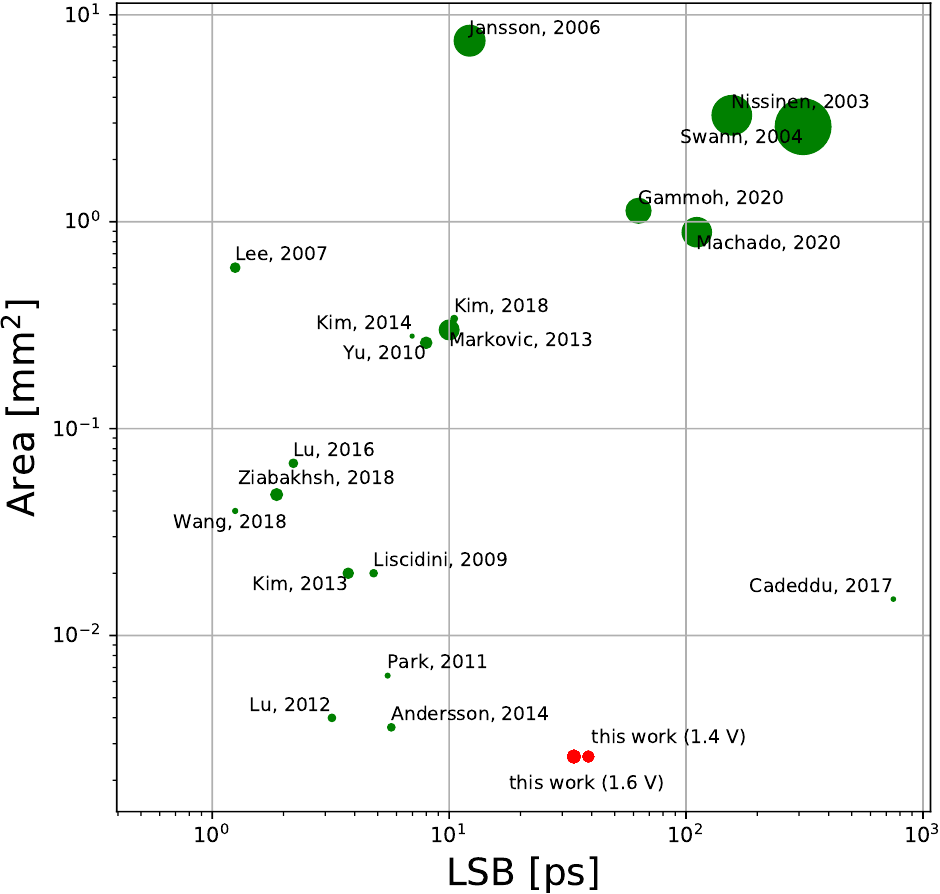}
\caption{Area and LSB of the presented TDC compared to the works of Table \ref{tdc_tab:post_layout_sim} and the ones reported in \cite{jansson2006cmos,wang2018reconfigurable,lee20079b, cadeddu2017time, gammoh2020linearity, machado2020technology, ziabakhsh2018second,kim201411, lu20123, yu201012, markovic2013high, kim20137, lu20162}. The size of the dots on the plot is proportional to the power consumption of the analyzed TDCs (logarithmic scale).}
\label{scatter_comparison}
\end{figure}

Table \ref{tdc_tab:post_layout_sim} offers a comparison between the TDC described in this paper and other works. As highlighted before, the main property of the presented TDC is the compactness and the simplicity of the PLL-less architecture which makes it the solution with the smallest area among all the cited works (for \cite{muntean2020blumino} the area is not reported). Solutions \cite{kim2018two} \cite{andersson2014vernier} \cite{park2011cyclic} and \cite{liscidini2009time} are characterized by smaller power consumption and LSB but they have been developed in a more advanced technological node and, as explained in Section \ref{sec:introduction}, the complexity and/or the limited maximum measurable time interval make them more difficult to be integrated in large pixel detector chips. The non-linearities of the presented architecture are comparable with the other works (only solutions \cite{swann2004100} and \cite{nissinen2003cmos} have significantly better values of DNL and INL but their power consumption is one or two orders of magnitude higher than the one of the PLL-less TDC). The performance of the converter proposed in this paper is compared to some of the works reported in Table \ref{tdc_tab:post_layout_sim} and in Figure \ref{scatter_comparison}. Even this plot highlights the compactness of our architecture compared to others with similar performance in terms of resolution and power consumption.

\section{Conclusion}

A RO-based TDC was developed to be integrated in pixel detectors for HEP and medical imaging applications. Simulations and measurements show a LSB of 33.6 ps (or 38.7 ps for lower supply) and a DNL$\leq$1.3 LSB. Two models were developed for the analysis of the proposed solution architecture and to demonstrate that the integration of the buffers into the feedforward paths is useful to reduce the impact of their mismatch on the linearity of the system. This solution does not add any complexity to a standard multi-path architecture since it only requires the buffers to drive the input of other delay cells other than the external loads. For this reason, this simple modification in the architecture of the system can be applied to any multi-path RO-based TDC in various technologies. The PLL-less event-by-event calibration system, the small power consumption and the compact area allow an easier integration of a large number of converters in pixel detector chips, a crucial characteristic for the above-mentioned applications.

\acknowledgments
The authors wish to thank the technical staff and the engineering team of the Physics Department at University of Geneva for their support in the preparation of the boards and the test setup. The authors would also like to thank A. Muntean from AQUA laboratory at EPFL for the productive and useful scientific discussions that helped the design process.


\bibliographystyle{unsrt}
\bibliography{mybib}







\end{document}